\documentclass{aa} 

\usepackage{txfonts}
\usepackage[usenames]{color}
\usepackage{natbib}
\bibpunct{(}{)}{;}{a}{}{,} 
\usepackage{graphicx}
\usepackage{float}
\usepackage{hyperref}

\def\setid#1,v #2 #3/#4/#5 #6 #7 #8.{(\textit{CVS version #2, checked in by #7 on #3-#4-#5, #6 UT})}
\definecolor{SanjColor}{rgb}{1,0,1}

\newcommand{\FeI}{\ion{Fe}{i}}

\newcommand{\kms}{\,km\,s$^{-1}$}
\newcommand{\ms}{\,m\,s$^{-1}$}
\newcommand{\eg}{e.g.}

\begin{document}

\titlerunning{Global properties of a sunspot}
\authorrunning{Tiwari, et al.}

\title{Depth-dependent global properties of a sunspot observed by Hinode (using the Solar Optical Telescope/Spectropolarimeter)}

\author{Sanjiv K. Tiwari$^{1,2}$, Michiel van Noort$^1$, Sami K. Solanki$^{1,3}$, Andreas Lagg$^1$}
\institute{$^1$Max-Planck-Institut f\"{u}r Sonnensystemforschung,
 Justus-von-Liebig-Weg 3, 37077 G\"{o}ttingen, Germany.\\
$^2$NASA Marshall Space Flight Center, ZP 13, Huntsville, AL 35812, USA.\\
$^3$School of Space Research, Kyung Hee University, Yongin, Gyeonggi 446-701, Republic of Korea.\\
\email{[tiwari;vannoort;solanki;lagg]@mps.mpg.de}}
\offprints{S.~K. Tiwari \email{sanjiv.k.tiwari@nasa.gov}}

\abstract
{For the past two decades, the three-dimensional structure of sunspots has been studied extensively. A recent improvement in the Stokes inversion technique prompts us to revisit the depth-dependent properties of sunspots.}
{In the present work, we aim to investigate the global depth-dependent thermal, velocity, and magnetic properties of a sunspot, as well as the interconnection between various local properties.}
{We analysed high-quality Stokes profiles of the disk-centred, regular, leading sunspot of NOAA AR 10933, acquired by the Solar Optical Telescope/Spectropolarimeter (SOT/SP) on board the Hinode spacecraft. To obtain depth-dependent stratification of the physical parameters, we  used the recently developed, spatially coupled version of the SPINOR inversion code.}
{First, we study the azimuthally averaged physical parameters of the sunspot. We find that the vertical temperature gradient in the lower- to mid-photosphere is at its weakest in the umbra, while it is considerably stronger in the penumbra, and  stronger still in the spot's surroundings. 
The azimuthally averaged field becomes more horizontal with radial distance from the centre of the spot, but more vertical with height. At continuum optical depth unity, the line-of-sight velocity shows an average upflow of $\sim$ 300 ms$^{-1}$ in the inner penumbra and an average downflow of $\sim$ 1300 ms$^{-1}$ in the outer penumbra. The downflow continues outside the visible penumbral boundary. The sunspot shows, at most, a moderate negative twist of $< 5^\circ$ at $\log(\tau) = 0$, which increases with height. The sunspot umbra and the spines of the penumbra show considerable similarity with regard to their physical properties, albeit with some quantitative differences (weaker, somewhat more horizontal fields in spines, commensurate with their location being further away from the sunspot's core). The temperature shows a general anti-correlation with the field strength, with the exception of the heads of penumbral filaments, where a weak positive correlation is found. The dependence of the physical parameters on each other over the full sunspot shows a qualitative similarity to that of a standard penumbral filament and its surrounding spines.}
{The large-scale variation in the physical parameters of a sunspot at various optical depths is presented. Our results suggest that the spines in the penumbra are basically the outward extension of the umbra. The spines and the penumbral filaments, together, are  the basic elements that form a sunspot penumbra.}

\keywords{Sun: magnetic fields -- Sun: photosphere -- Sun: sunspots}
\maketitle

\section{Introduction}
High-resolution observations of sunspots, which are the dark features on the solar surface, reveal that they contain small-scale, dynamically evolving structures, such as umbral dots \citep{dani64,riet08,riet13}, light bridges \citep{sobo97,rimm08,shim11,lagg14}, spines \citep{lites93}, penumbral filaments \citep[and references therein]{tiw13}, and concentrated strong downflows at the spot's periphery \citep{van13}. In spite of the fast dynamical evolution of these elementary features \citep[$\sim$ 10 minutes to 4 hours, see for example,][]{sobo97b,sola03a}, the sunspots are usually long-lasting, existing for a few  days to several months, and only evolve  slowly. To understand this relative stability, it is essential to study the global properties of sunspots, in addition to exploring their fine-scale structure. The global properties of sunspots are also of interest as constraints for global spot models, for studies of active region magnetic fields and associated activity, for investigations of solar irradiance variations, and as proxies for starspots. 

In the past two decades, considerable progress has been made in understanding the three-dimensional (3D) sunspot structure in the solar atmosphere, both theoretically \citep{remp09a,remp09b,remp11,remp12} and observationally \citep{sola92,lites93,titl93,stan97,west98,west01,west01a,sola03,shibu03,borr11}. Many of the depth-dependent global properties of the magnetic, thermal, and velocity fields in sunspots have been verified by those researchers. For example, within sunspots the magnetic field strength increases, whereas field inclination generally decreases with depth, the temperature increases with both depth and radius, and the line-of-sight (LOS) velocity shows net upflows and downflows in the inner and outer penumbra, respectively.

When looked at in detail, the more vertical fields in the penumbra (spines, according to \citealt{lites93}) were found to be stronger than their surroundings. However, their thermal structure remains controversial. \cite{wieh00} associated dark penumbral features with the stronger and more horizontal fields. Like \cite{west01}, \cite{lang05} found the stronger penumbral magnetic field to be more vertical and brighter. \cite{borr11} found more horizontal fields (intra-spines) to be weaker and brighter in the inner penumbra and darker in the outer penumbra \cite[see][for a detailed review of earlier literature]{sola03}.
These kinds of controversies were recently addressed by \cite{tiw13}, who show that the heads of sunspot penumbral filaments are brighter and contain vertical magnetic field, whereas their tails are darker and contain oppositely directed, strong, vertical magnetic fields. Only the darker regions that are less inclined to the vertical fields of the same polarity as the umbra, can be associated with the spines. It has  also been confirmed that the magnetic field in the spines wraps around the intra-spines in higher layers, as anticipated by the models of \cite{sola93} and \cite{spru06}, and reported by \cite{borr08}.

However, many issues remain to be resolved. As an example, the magnetic canopy structure, which is a result of the expansion of the magnetic field, has not been unambiguously observed and understood. \cite{west01,reza06} and \cite{borr11} find that the canopy starts well within the sunspot penumbra and continues outside of it, whereas \cite{shibu03} do not find a magnetic canopy structure anywhere. \cite{balt08} find the canopy only outside the penumbral boundary, in contrast to the above-mentioned researchers, but in agreement with, for example, \cite{giov80,giov82,sola92,sola94,sola99} and \cite{adam93}. 

Similarly, the relationship between the temperature and magnetic field strength over the full sunspot is not fully understood \citep{sola03}. A non-linear relationship between temperature (and/or intensity) and magnetic field is observed by \cite{sola93a,shibu04}, which   agrees with the anti-correlation found by other researchers, e.g. \cite{kopp92,pill93} and \cite{stan97}. The above authors attributed the non-linearity in the (penumbral) lower temperature part of their scatter plots to the outer penumbral features. However, the different parts of the non-linearity in the scatter plots between field strength $B$ and temperature $T$, e.g. a nearly constant temperature (of $\sim$5800$-$5900 K) for a range of magnetic field strength (0$-$2500 G) found by \cite{west01}, remains unexplained. 

The magnetic field strength and inclination are anti-correlated, with more vertical fields having higher field strength \citep[e.g.][]{stan97,west01}. This relationship mirrors the large-scale structure of sunspots. On smaller scales we expect a relationship that is affected by the local magnetoconvective structure, with strong opposite polarity fields at the tails of the penumbral filaments \citep{tiw13}; see also \cite{scha13};  \cite{ruiz13}.

The source of the Evershed flow \citep{ever09} is near the heads of the penumbral filaments, with the gas cooling down as it travels the bulk of the filaments and sinks as cool material at the tails of the filaments \citep{tiw13}. Based on this, a clear, systematic relationship between temperature and LOS velocity was found by \cite{tiw13} for a standard penumbral filament, where the upflows are hotter and the downflows are cooler. As an extension, in the present work, we perform this study on a full sunspot to investigate how this pattern, together with the cool spines, affects the structure of the full sunspot.

In spite of the large number of works on global sunspot properties \citep[see references above, and for example,][]{lites93,west01,west01a,sola03,shibu03,trit04,bell04,borr11}, a description of these properties in terms of elementary, small-scale structures, such as umbral dots, spines, and penumbral filaments, has not been presented to date.

Recent advances in the inversion of the high-quality spectro-polarimetric observations from Solar Optical Telescope/Spectropolarimeter (SOT/SP) provide us with an opportunity to take a fresh look at the above-mentioned issues in detail. 
In the present paper, we analyse the results of a depth-dependent inversion of a full sunspot and its surrounding region. The same inverted results are analysed, as was done earlier by \cite{tiw13}. While we concentrated mainly on the internal structure of penumbral filaments in that paper, here we focus on the global and partly local properties of the inverted sunspot. 

This paper is organised as follows: Details of the observations and the inversion process are given in the next section. In Section 3, we present the depth-dependent global thermal, velocity, and magnetic properties of the sunspot. The mutual dependence of physical parameters and inferred substructures of the sunspot are given in Section 4. In Section 5, we discuss our results, and finally, in Section 6, we present our conclusions.

\section{Observations and inversion}

We used a high spatial resolution spectro-polarimetric scan of the leading sunspot of the active region (AR), NOAA 10933, recorded with the SOT/SP (\citealt{tsun08,suem08,ichi08,shim08}) onboard the Hinode satellite \citep{kosu07}. The positive polarity spot of the  NOAA AR 10933 that we use in the present study was observed close to the solar disk centre ($\mu$=0.99) on  5$^{}$  January 2007 from 12:36 to 13:10 UT. The atomic lines contained in the data are the Fe {\footnotesize I} lines at 6301.5 and 6302.5 \AA. The SP scans are acquired in normal mode of SOT, with a spatial sampling of $0.16$ \arcsec pixel$^{-1}$.

\begin{figure}[htp]
\centering
\includegraphics[width=\columnwidth]{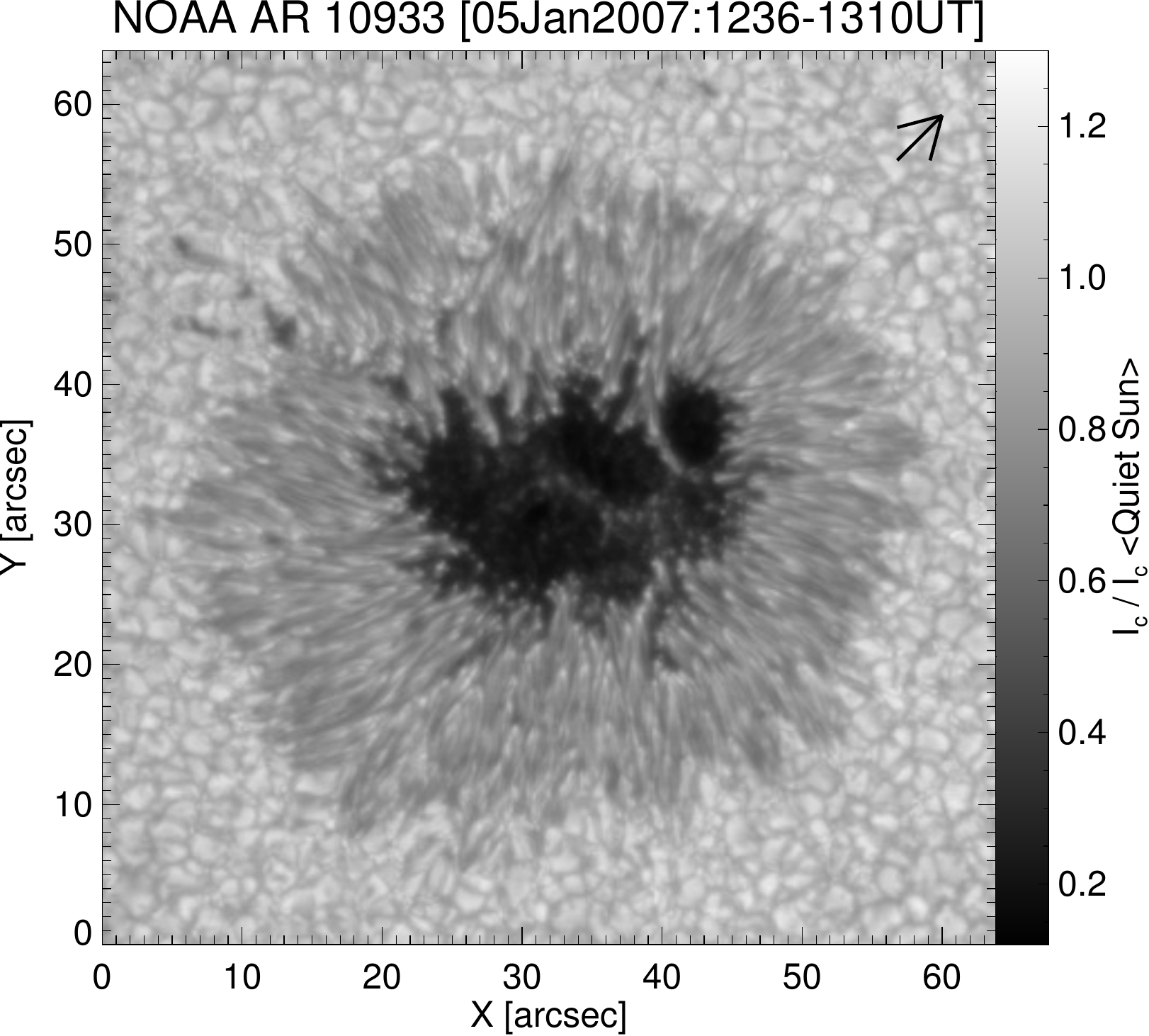}
\caption{Continuum intensity map of the sunspot NOAA AR 10933, observed by Hinode in normal scan mode of SOT/SP on 5 January 2007 between 12:36 and 13:10 UT. The arrow indicates the direction of the solar disk centre.}
\label{cont_obs}
\end{figure}

\begin{figure*}[p]
 \centering
 \includegraphics[width=0.99\textwidth]{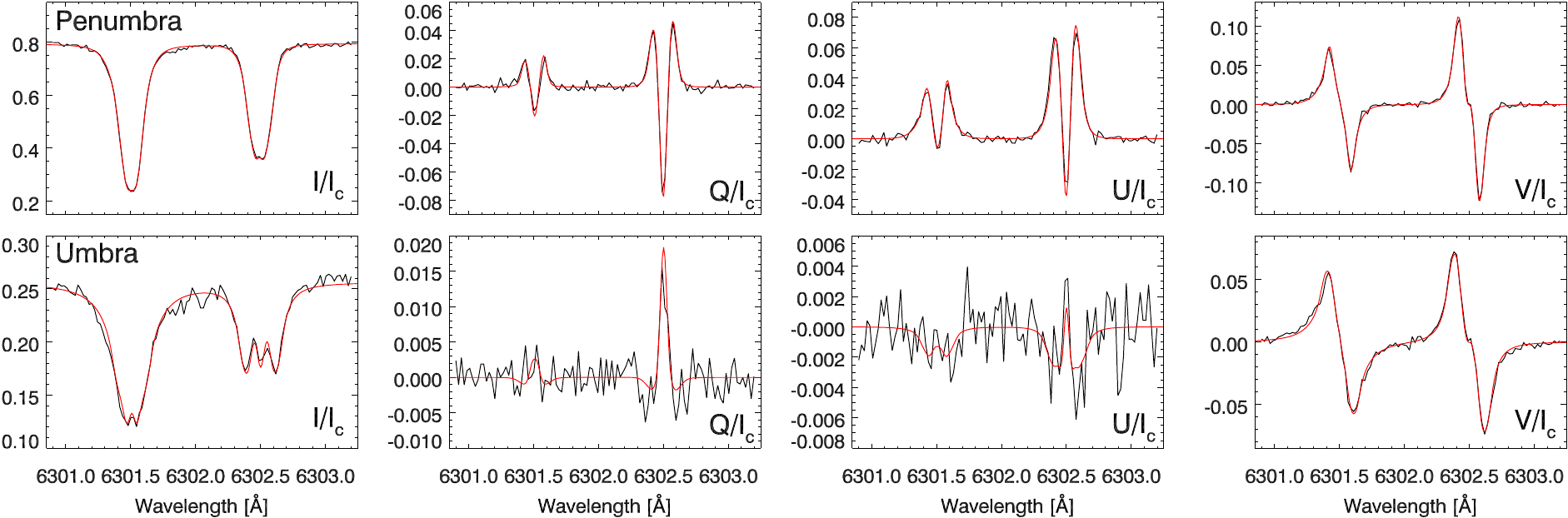}
 \caption{Examples of observed (black) and best-fit (red) Stokes profiles: I, Q, U, and V. Upper row displays Stokes profiles for a typical pixel in the sunspot penumbra (location: X48\arcsec, Y41.6\arcsec). Lower row contains Stokes profiles for one of the darkest pixels in the sunspot umbra (location: X36\arcsec,Y32\arcsec). }
 \label{profiles}
\end{figure*}

We perform the calibration of the Hinode (SOT/SP) data by using the standard SP\_PREP routine, which is available in the SOLAR software package. The SP\_PREP routine  computes the thermal shifts in the spectral and slit dimensions first, and then applies the drift corrections for calibrating the data from Levels 0 to 1 \citep{lites13}.

The observed continuum intensity map of the inverted region is shown in Fig. \ref{cont_obs}. The spot has an average diameter of about 50\arcsec, and different aspects of it have already been  studied in a number of publications: see for example, \cite{kubo08,tiw09b,fran09,venk09,venk10,kats10,fran11,borr11,tiw12,riet13,van13,tiw13}.

\subsection{Inversion}
\label{inversion}

We employ the output of the SPINOR inversion code \citep{frut00,frut00b} in the spatially coupled mode \citep{van12,van13} used by \cite{tiw13} to invert the calibrated Stokes profiles. The SPINOR code builds on the STOPRO line synthesis routines by \cite{sola87} and solves the radiative transfer equations numerically, under the assumption of local thermodynamical equilibrium (LTE). The spatially coupled version of the SPINOR inversion code employs a modified Levenberg--Marquardt (L-M) algorithm for the optimisation of a simplified, height-dependent atmospheric model, simultaneously over an extended field of view, taking the spatial degradation caused by the telescope into account. The required spectro-polarimetric derivatives, with respect to the fitted atmospheric structure required by the L-M algorithm, are calculated using response functions \citep{landi77,ruiz92}. To resolve the maximum amount of spatial structure, a Fourier interpolation of the data was made to allow the inversion to be carried out at an increases spatial sampling by a factor of two \citep{van13}. Thus, the pixel size for resultant physical parameters used in the study is 0.08\arcsec. 

The vertical structure of the temperature $T$, magnetic field strength $B$, inclination $\gamma$, azimuth $\chi$, LOS velocity $v_{\rm LOS}$, and microturbulent velocity $v_{\rm mic}$ are approximated using bicubic splines, controlled by three nodes, placed at log($\tau$) = $-$2.5, $-$0.9, and 0, where $\tau$ is the continuum optical depth at $\lambda$ = 6302.5 \AA. 

To help with the convergence of the solution, for every ten iterations, we re-initialised all slowly converging pixels   by using their nearest neighbours, and the whole solution was then smoothed over by convolution with a Gaussian smoothing function. This significantly speeds up convergence if appropriate selection criteria for re-initialization are used. Although this procedure can create a preference for a spatially smooth solution, it clearly results in a significant increase in the fit quality, i.e. a reduction in the difference between square of the data and the fitted profiles, i.e. a significantly lower $\chi^2$. While, as in all inversions, the adopted solution may or may not represent the global optimum, it does appear to represent a workable optimum, since additional attempts to disturb and reconverge the solution did not significantly increase the accuracy of the fit, and resulted in a very similar solution.

\begin{figure*}[htp]
 \centering
 \includegraphics[width=\textwidth]{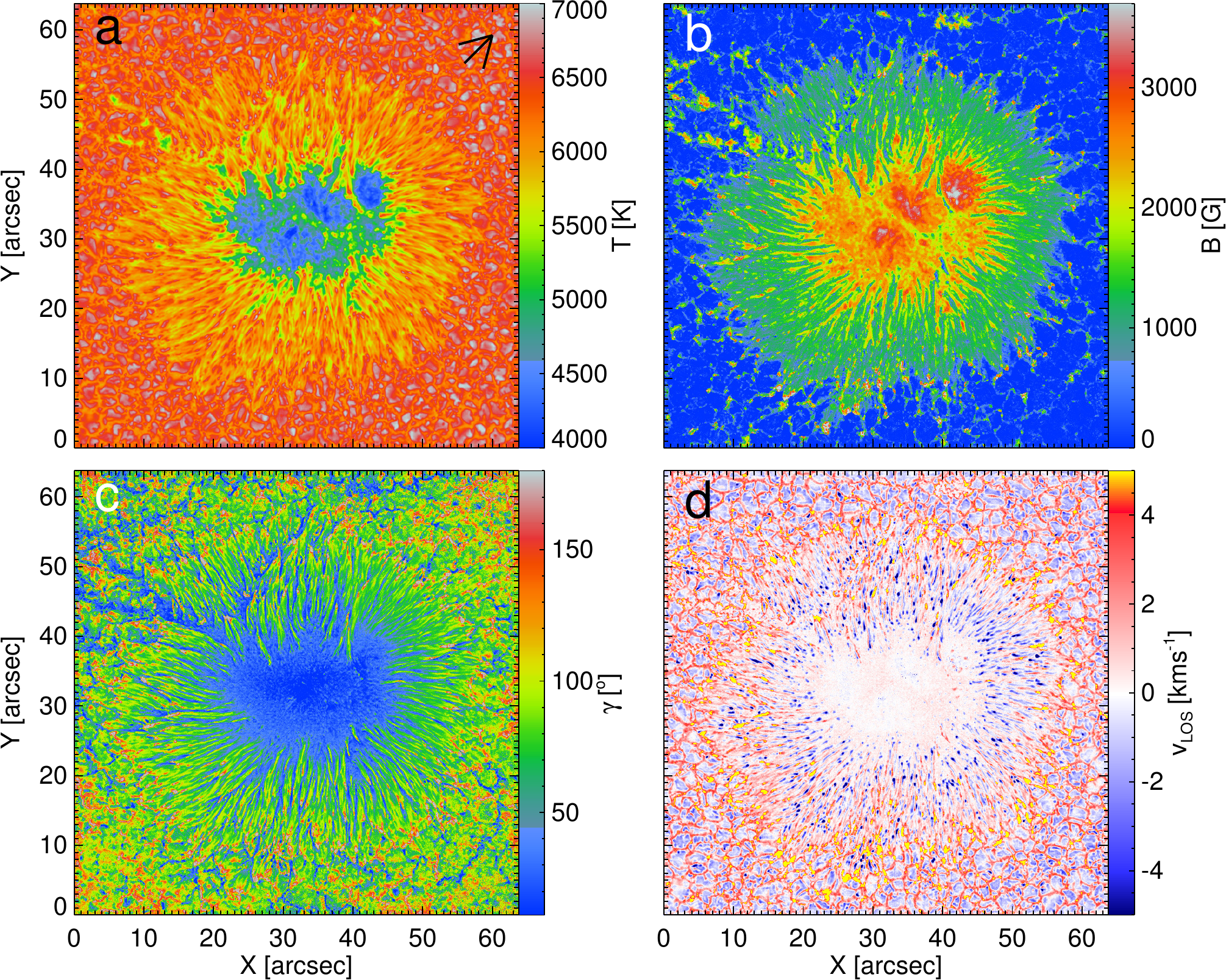}
 \caption{Maps of selected physical parameters (at log$(\tau)$ = 0) obtained from the inversion: (a) temperature $T$, (b) magnetic field strength $B$, (c) magnetic field inclination $\gamma$, (d) line-of-sight velocity $v_{\rm LOS}$. The arrow in the temperature map points to the solar disk centre. Except for field inclination $\gamma$, thresholds have been used for minimum and maximum values of the colour bars. The extreme values returned by the inversion are 3853 to 7173 K for $T$, 10 to 4000 G for $B$, and $-$11.5 to 18.7 \kms\ for $v_{\rm LOS}$.}
 \label{full_maps}
\end{figure*}

Examples of observed and fitted Stokes profiles for a typical pixel, each one sample taken from penumbra and umbra, are shown in Fig \ref{profiles}. The atomic parameters used in the inversion are 6302.4936: log(gf)=$-$1.203, E$_I$=3.6866 eV, and 6301.5012: log(gf)=$-$0.745, E$_I$=3.6539 eV. Collisional broadening was calculated using the ABO tables \citep{bark98} and \FeI\ abundance was set to 7.50. No molecular lines are included in the inversion since it makes the inversion process extremely slow. This sunspot was not particularly cool, meaning the inversion does not seem  to have been severely affected by contamination through molecular lines (see, e.g. the  fits for one of the darkest pixels in the sunspot umbra in Fig.~\ref{profiles}).

A test was performed by \cite{van13} to establish the robustness of the inversion. This involved changing the node positions and adding different levels of noise in the observed and convolved profiles and resulted in very similar inverted quantities, which suggests that the inversion result is reliable. Moreover, \cite{van12,van13} find good agreement between the inversion results and 3D radiation magnetohydrodynamics (MHD) simulations with the MURAM code \citep{vogl05} for the velocities and fields, providing further support to the reliability of our inversion process. This result is supported by the fact that, for most pixels, the response functions are non-zero for all three chosen nodes, particularly for the one at log($\tau$) = 0.

However, establishing the error in the fitted atmospheric parameters is a tedious task because of the spatial coupling of the parameters. A change introduced in a parameter at one pixel will cause the code to adjust the fit of that and other parameters in the neighbouring pixels. Therefore, determining the error in the fitted parameters would require a perturbation of each parameter at every pixel, followed by a full reconversion of the inversion in all the other pixels. This  would demand a prohibitively large amount of computational power.

The wavelength calibration of the data was carried out by requiring that the umbra of the sunspot, excluding the umbral dots, be on average at rest at $\tau=1$. This resulted in a small systematic correction of the LOS velocity, $v_{\rm LOS}$, of 100 \ms, which was subtracted from the inversion results presented here.

The 180$^\circ$ azimuthal ambiguity was resolved after the inversion by using the minimum energy method \citep{metc94,leka09}.
In Fig. \ref{full_maps}, we display full maps of inversion results of some of the physical parameters (fitted temperature $T$, magnetic field strength $B$, field inclination $\gamma,$ and LOS $v_{\rm LOS}$) at log($\tau$) = 0. The physical parameters look smooth, in addition to having high contrast (e.g. compare the observed continuum intensity from Fig.~\ref{cont_obs} with the one obtained from the inversion shown in Fig.~\ref{contours_cont}, and discussed later). At log($\tau$) = $-$0.9, where the response of the spectral lines 6301.5 and 6302.5 \AA\ to most of the physical parameters is largest, the physical parameters become smoother than at log($\tau$) = 0.

\section{Global properties}
\label{global}
\begin{figure}[tp]
\centering
\includegraphics[width=\columnwidth]{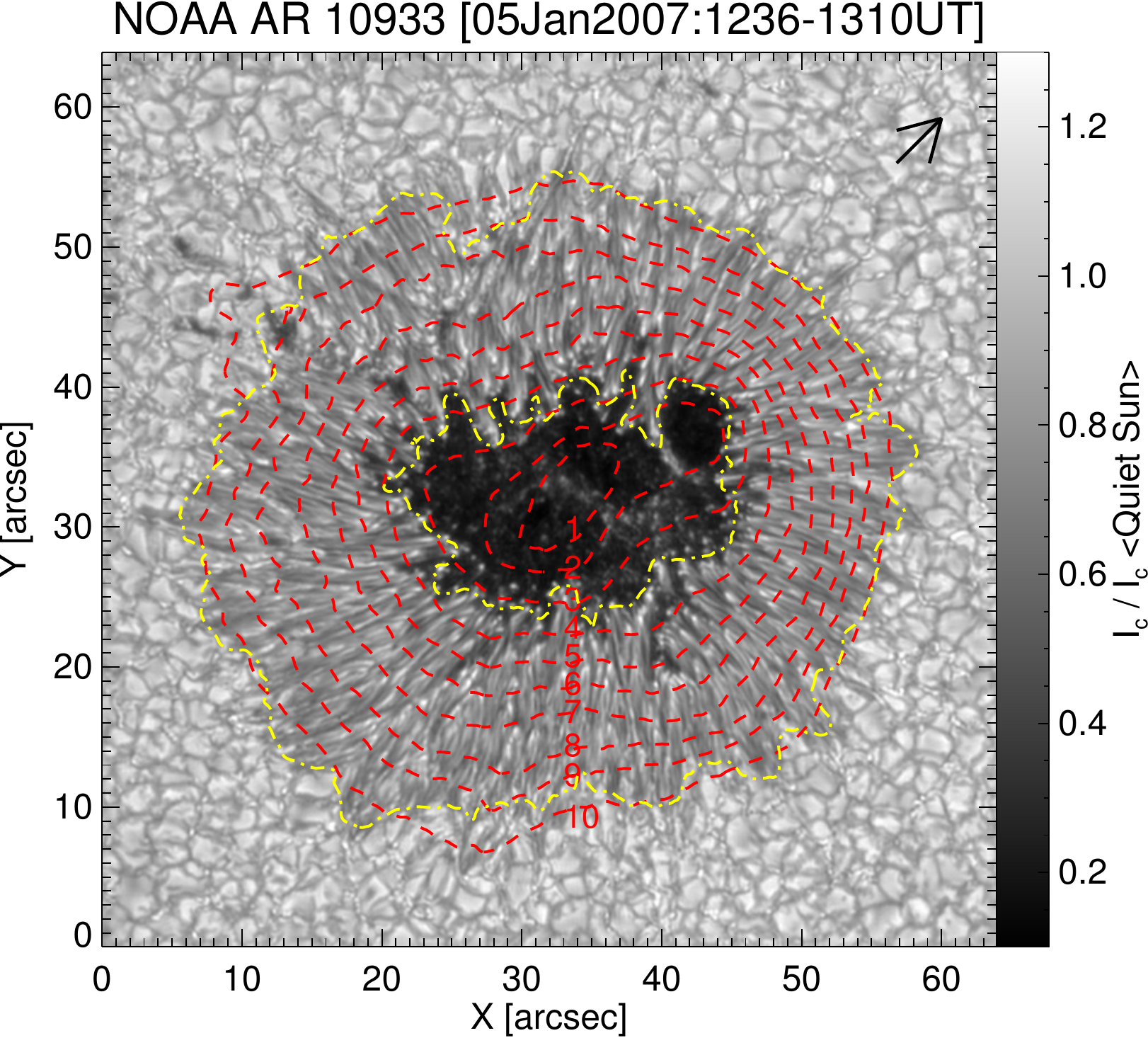}
\caption{Ten of the 21 contours of magnetic field strength over-plotted on the inverted continuum intensity map. Red contours are based on the smoothed magnetic field strength. Yellow contours are based on the intensity and represent the boundaries of the sunspot penumbra. The greyscale bar ranges from 0.1 to 1.3; the extreme values of $I_{c}/I_{c}$$<${\it Quiet Sun}$>$ returned from the inversion are 0.097 and 1.468. The black arrow on the upper right part of the image points towards the solar disk centre.}
\label{contours_cont}
\end{figure}

To investigate the global properties of the sunspot, we selected several contours of magnetic field strength of the spot, which was first smoothed by a factor of about twenty (20$\times$20 pixels) to get smooth contours. Outer contours required smoothing by even larger factors (up to 50) because there are more fluctuations along the azimuthal path that is present there. Ten contours out of a total of 21 are shown in red in Fig.~\ref{contours_cont}. The inner and outer penumbral boundaries are determined on the basis of intensity and are indicated by the yellow contours in Fig. \ref{contours_cont}.

We note that both the outermost red contour and the outer yellow one differ quite significantly in some places. This is partly due to the different amount of smoothing applied to the field strength and continuum intensity prior to determining the contours, but partly also to a mismatch between field strength and intensity near the penumbral boundary. We  looked carefully at these regions and found that many of them are the tails of penumbral filaments with high-speed downflows, and contain very strong fields with a polarity opposite to that of the sunspot umbra \citep{van13}. In most of these regions the tails of several penumbral filaments seem to converge. The most prominent excursion of the outermost red contour beyond the yellow contour (X $\approx 10\arcsec$, Y $\approx$ 45\arcsec) is in the direction of a group of pores of the same polarity, i.e. in a direction in which the penumbra is strongly distorted.

\subsection{Thermal properties}
\begin{figure}[tp]
 \centering
 \includegraphics[width=\columnwidth]{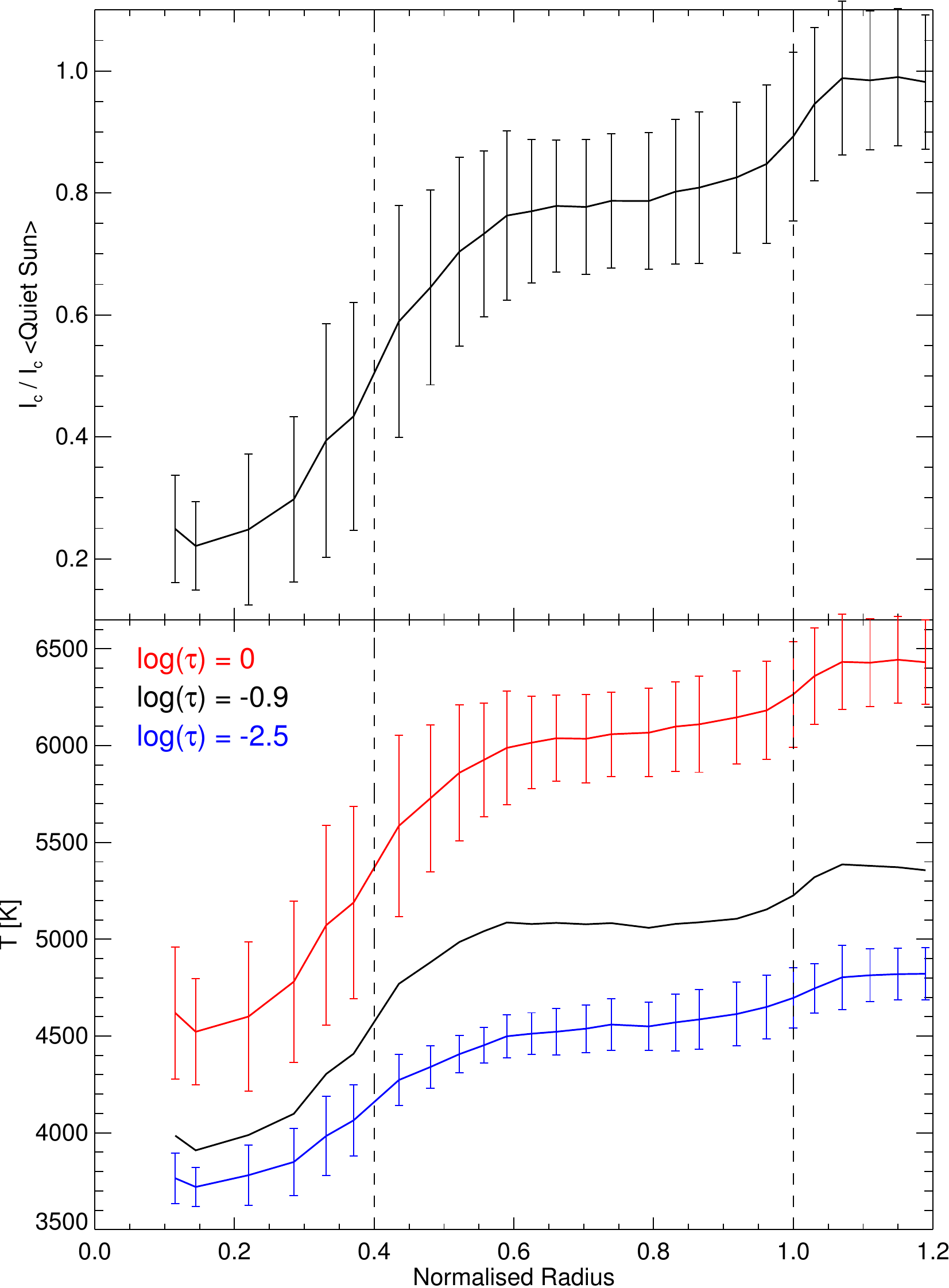}               
 \caption{Upper panel: Radial distribution of the azimuthally averaged continuum intensity normalised to averaged quiet Sun. The error bars represent the RMS variation along the azimuthal path of the contours. Lower panel: Radial dependence of the azimuthal averages of temperature at three optical depth positions. The two vertical dashed lines represent average radial positions of the boundaries of the sunspot penumbra.}
 \label{cont_int_rad}
\end{figure}

To explore the variation in the continuum intensity and temperature along the radius of the sunspot, we averaged these parameters azimuthally along the given contours (i.e. at the three optical depth  positions for temperature that correspond to the nodes used for the vertical spline interpolation of the atmosphere). The mean values, along with the root mean square (RMS) variation of the sample along the path of the contours (not to be confused with the standard error to the mean), are plotted in Fig. \ref{cont_int_rad}. The RMS variations are quite significant, indicating the presence of fine-scale structures along the path of the contours. The contours are based on the magnetic field strength and are not equidistant. Because of this, the corresponding error bars are also not equidistant.

The brightness of the umbra increases with radial distance from the centre of the spot. We find that the umbra  has an azimuthally averaged intensity of 20-40\% of the quiet Sun in the continuum at 6302.0 \AA, in agreement with earlier findings \citep[e.g.][]{west01}. The darkest part of the umbra has continuum intensities as low as 10\% of the quiet Sun. The penumbra has a brightness of 45-85\% of that of the averaged quiet Sun. However, these values depend on the criterion used to determine the umbra-penumbra boundary and on the wavelength. (See, for example, \cite{shibu03} for infrared wavelengths). They also depend on the size of sunspots \citep[and references therein]{shibu07}, whereas the purported dependence on the phase of solar cycle remains inconclusive \citep{albr78,albr81,albr84,malt86,nort04,penn06,shibu07}. 

The behaviour of the azimuthally averaged temperature at the deepest node $\log(\tau)=0$, shown in the lower panel of Fig.~\ref{cont_int_rad}, closely mimics that of the continuum intensity, displayed in the upper panel of the same figure. This is  to be expected for an LTE inversion result. The temperature drops more rapidly from penumbra to umbra in the deeper layers than higher up,  suggesting a lower vertical  temperature gradient in the spot. Whereas in the penumbra this vertical temperature gradient is only slightly weaker than that in the quiet Sun, it is much weaker in the umbra. The latter result is in good agreement with the literature \citep[see, e.g.  ][]{schr71,malt86,coll87,west01,sola03,trit04}.

The temperature shows the largest fluctuations at $\log(\tau)=0$. Since the temperature is very well constrained in this layer, largely by the continuum intensity, such strong fluctuations are an intrinsic property of the sunspot atmosphere, and are caused by the fine-scale structures in it.

\subsection{Velocity stratification}
Our sunspot was located almost at the centre of the solar disk (within 5$^\circ$), allowing us to treat the LOS velocity as the vertical velocity, and to ignore projection effects. We find that the deepest layer in the inner penumbra is dominated by upflows (see Fig.~\ref{full_maps}d), in agreement with earlier findings. This is as expected from the dominating presence of heads of penumbral filaments, which are found in this location of
the spot, and which  contain strong upflows \citep{tiw13}. At the outer boundary of the sunspot penumbra, strong downflows dominate in the deepest layer, again as expected from the known concentrations  of the tails of penumbral filaments with strong downflows that are found at this location \citep{van13,tiw13}. This large-scale radial structure of flows was seen earlier by \cite{rimm95,schl00,west01a,trit04,rimm06,ichi07,sanc07} and \cite{fran09}. 

\begin{figure}[htp]
 \centering
 \includegraphics[width=\columnwidth]{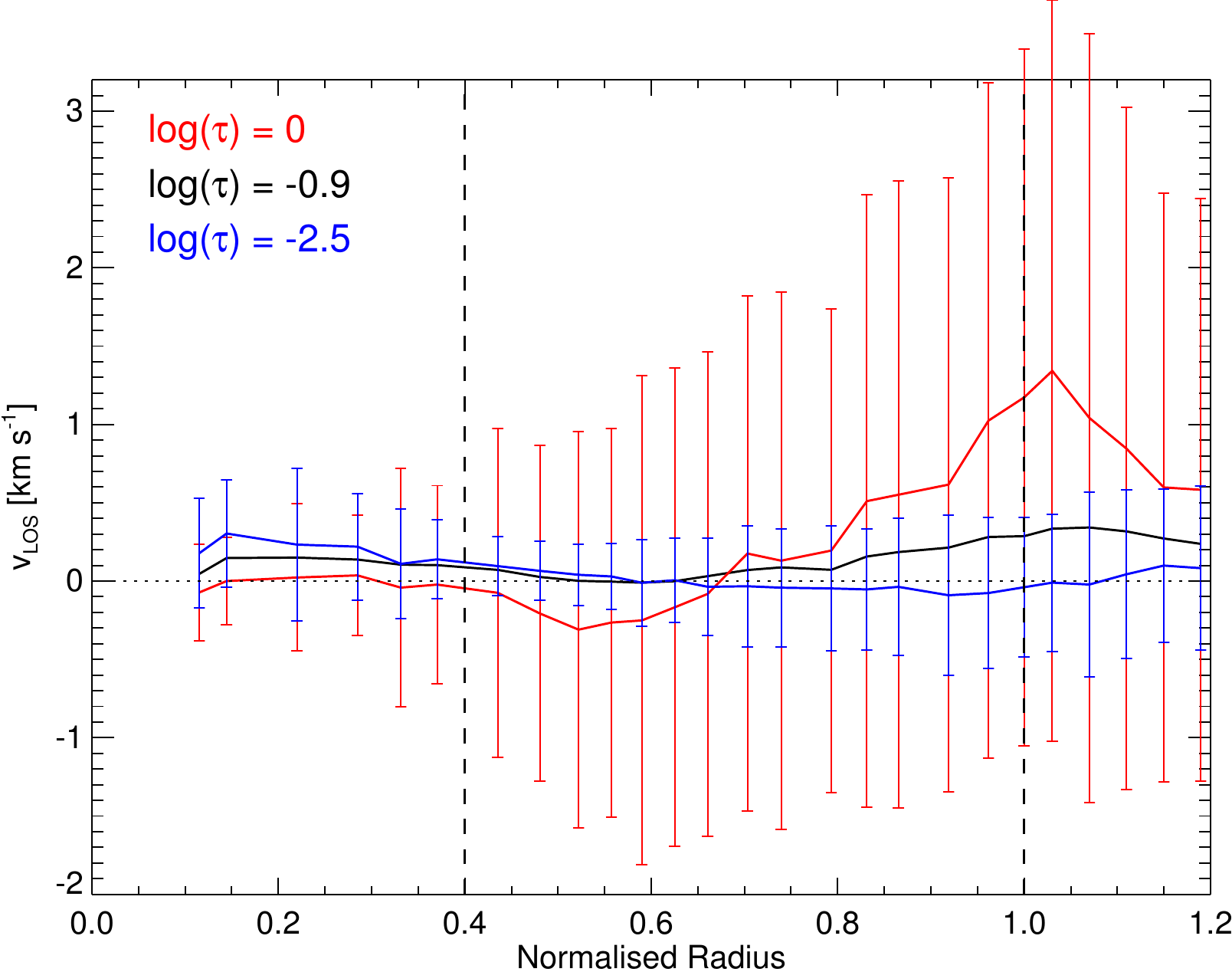}
 \caption{Azimuthal averages of $v_{\rm LOS}$ at three optical depths plotted vs normalised radial distance from centre of the spot (solid lines). The RMS variations along each azimuthal contour are indicated for log($\tau$) = 0 and $-2.5$ by the red and blue vertical bars, respectively. The two vertical dashed lines have the same meaning as in Fig. \ref{cont_int_rad}.}
 \label{vel_temp_rad}
\end{figure}

In Fig.~\ref{vel_temp_rad}, we present the azimuthally averaged $v_{\rm LOS}$ at the three depth node positions vs radius. This plot clearly shows the domination of upflows in the inner penumbra with the average velocity in the deepest layer around $-$300 \ms\  and the even greater domination of downflows in the outer penumbra, where the average velocity reaches values as high as 1300 \ms\ just outside the sunspot boundary. The rapid increase in the RMS variations with depth is in good agreement with the results of other studies of the fine structure of penumbrae \citep[see for example][and references therein]{tiw13}. The increase in RMS fluctuations  is nearly linear, with radial distance up to the outer boundary of the spot, which indicates increased inhomogeneity in the vertical velocity.   

The average downflow in the deepest layer increases in strength up to a normalised radius of $\sim$ 1.05, outside the visible boundary of the sunspot. 
The increase of the velocity beyond the boundary of the spot is a result of some of the strongest downflowing regions  not being inside the contour that marks the sunspot boundary, as pointed out in the beginning of Section \ref{global}. Furthermore, the downflows continue well beyond the boundary of the sunspot, in agreement with \cite{boer92}.  

The average velocity in the umbra is zero at log($\tau$) = 0, which was set by the velocity calibration described in  Section \ref{inversion}. The average velocity in the higher layers of the umbra remains close to (but different from) zero throughout the sunspot. In particular, there is a systematic weak downflow in the umbra above $\log(\tau)=0$, which increases with height, although, as we discuss in Section \ref{dis_gp}, the reliability of this result remains uncertain.  

The average velocity in the penumbra is a significant downflow of about 350 \ms\ at $\log(\tau)=0$, as expected from the presence of strong downflows at the outer boundary of the penumbra. This average downflow decreases with height, reaching  zero at $\log(\tau)= -2.5$.

\subsection{Magnetic properties}
\subsubsection{Azimuthal averages of field strength $B$ and inclination $\gamma$}
\begin{figure}
 \centering
 \includegraphics[width=0.91\columnwidth]{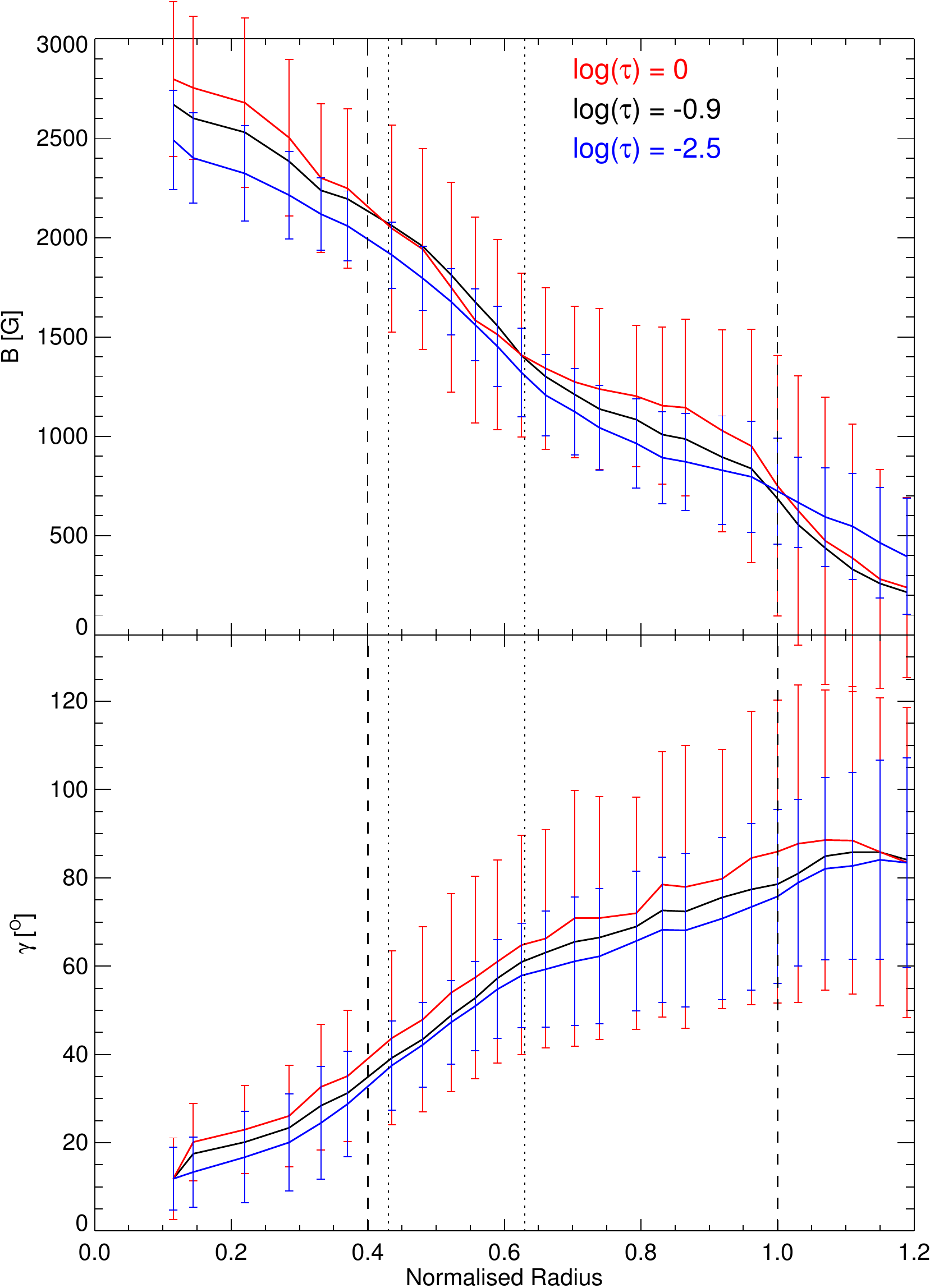}         
 \caption{Azimuthal averages of $B$ (upper panel) and $\gamma$ (lower panel) along the sunspot radius at three optical depths. Also shown are RMS variations for the uppermost and lowermost node positions (blue and red error bars, respectively). The two vertical dashed lines represent the inner and outer boundaries of the sunspot penumbra (on average). The two dotted lines outline the radius between which a positive field gradient is seen (i.e., field strength increasing upward).}
 \label{fld_incl_rad}
\end{figure}

Azimuthally averaged values of $B$ and $\gamma$, along with their RMS variations along the contours, are plotted in Fig.~\ref{fld_incl_rad}. The large RMS variations are again indicative of the presence of fine-scale structures along the path of the contours, although in the umbra the RMS variations tend to indicate the larger-scale inhomogeneity of $B$ and $\gamma$. The lower values of the error bars go below zero for the field strength for the contours outside the penumbral boundary because the distribution of the field strength along the path of these contours departs significantly from Gaussian.

We can see that the decrease in the average magnetic field strength is nearly linear with radial distance from the centre of the spot. The drop in field strength at log($\tau$) = 0, from 2800 G (in umbra) to 700 G (at outer penumbral boundary), is slightly larger than at the other two heights. Also noticeable is the substantial weakening of the field strength in the two lowest nodes outside the sunspot boundary.

The general trend toward a decrease in the field strength with increasing radius is similar to what is observed by \cite{west01,shibu03}, and \cite{borr11}, but the height dependence of $B$ in our plots, as discussed later, does not match  those of \cite{west01},  \cite{borr11}, or \cite{shibu03}. 

Fig.~\ref{fld_incl_rad} shows that there are two regions where the azimuthally averaged magnetic field strength increases with height: 1) just outside the visible boundary of the sunspot penumbra, and 2) in the region between radial positions 0.43 and 0.63. These are the locations of canopy structure and positive field gradient, respectively, which we describe in the next section (\ref{fgc}).   

The azimuthally averaged field inclination increases with radius, in agreement with the behaviour of an expanding flux tube. The averaged magnetic field inclination consistently exhibits more vertical fields in the higher layers in the umbra and penumbra. The magnitude of this vertical gradient, e.g. 5-10$^\circ$ between the lower and middle photosphere, is sizable. Outside the sunspot, however, the field in the deepest layer becomes more vertical and azimuthal averages of inclinations at all  three heights are nearly the same.

The fluctuations in both the field strength and inclination are largest in the deepest layer, and those in the inclination increase towards the outer boundary of the sunspot.

\subsubsection{Field gradients and canopy structure}
\label{fgc}

To estimate the magnetic field gradient, we need to obtain the field strength on a geometrical height (z-) scale, whereas the inversion code delivers the stratification of each physical parameter on a continuum optical depth ($\tau$-) scale. The transformation of optical depth scale to geometrical height is not straightforward because of the unknown optical corrugation of the surface over sunspots. We therefore carried out this transformation by assuming hydrostatic equilibrium. We are well aware that this is a gross simplification since it neglects the influence of the magnetic field (see also \cite{pusc10}).

\begin{figure}[h]
 \centering
 \includegraphics[width=\columnwidth]{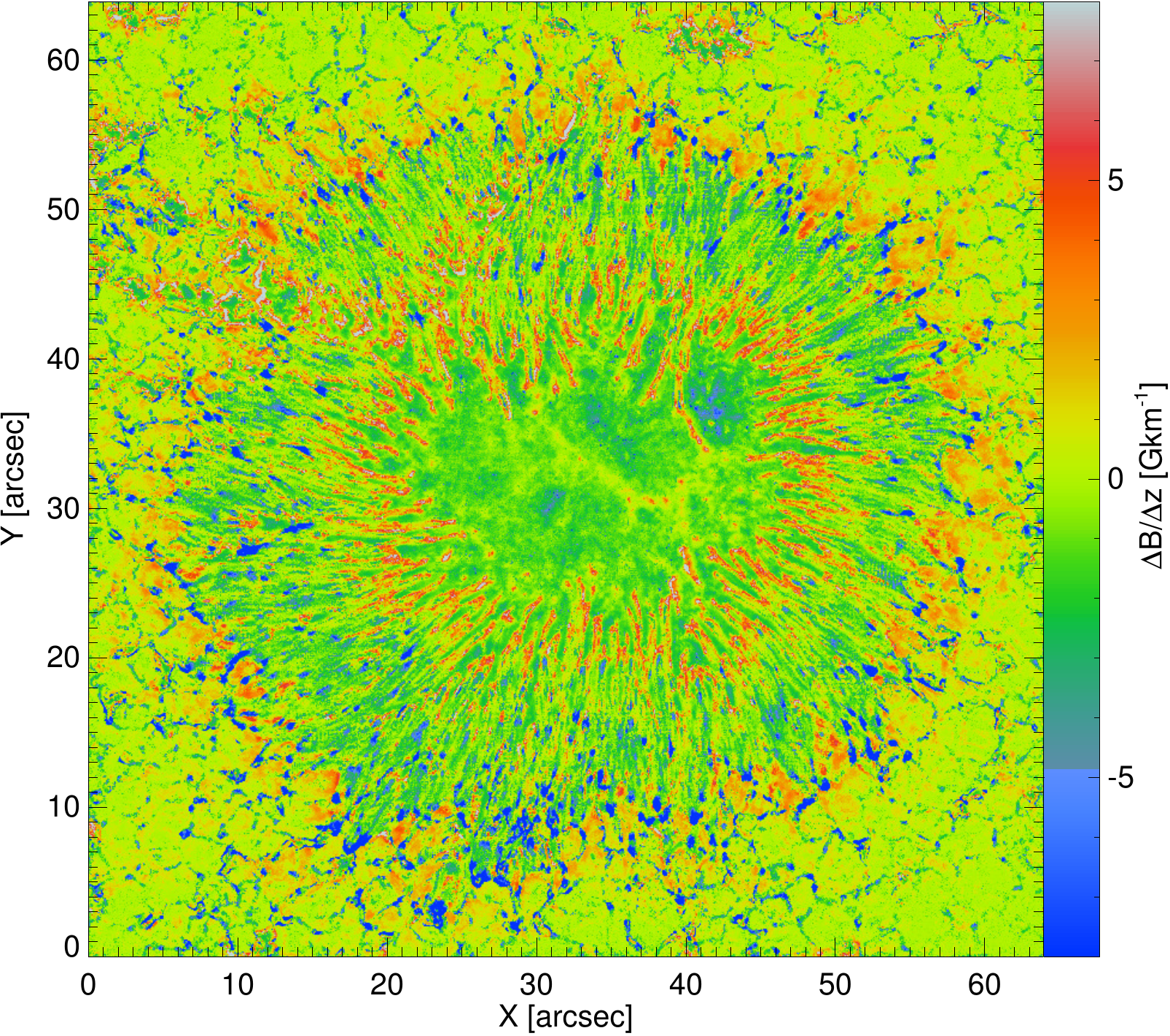}
 \caption{Gradient of magnetic field computed between the geometrical height corresponding to the two deepest nodes (log($\tau$) = $-$0.9, 0). A positive field gradient represents field strength increasing with height.}
 \label{gradb}
\end{figure}

The average vertical field gradient $\partial B/\partial z$ in the sunspot umbra near the deepest node, where the highest field strength is about 4~000 G, was determined to be $-$1.4 ~G km$^{-1}$. The gradient becomes weaker with height and reaches a value of $-$0.95 G km$^{-1}$ at log($\tau$) = $-$2.5. The magnitude of the field gradient found for our sunspot is equivalent to that found in 3D magnetohydrodynamic (MHD) simulations ($-$1.5 G km$^{-1}$ for a sunspot whose strongest magnetic field is close to 4~000 G at $\tau$ = 1: M. Rempel, 2013, private communication).

The azimuthally averaged magnetic field shows a positive gradient (opposite to that seen in the umbra) in the inner-middle part of the penumbra (roughly between radius = 0.43 and 0.63), and just  outside the sunspot (see upper panel of Fig.~\ref{fld_incl_rad}). The positive gradient observed outside the sunspot penumbra can clearly be associated with the magnetic canopy structure \citep{giov80,giov82,sola92,sola93b,sola94,sola99}, whereas the inverse gradient seen in the inner-middle part of the penumbra's deepest layers must have a different source. Figure~\ref{gradb} clearly reveals the locations of the positive field strength gradient present in the inner-middle penumbra and at the outer penumbral boundary. Such an inverse gradient is also seen in Vacuum Tower Telescope \cite[VTT:][]{schr85} data by \cite{josh15}, and Joshi (2014), who analysed this particular property of the sunspots in detail. They also found similar positive gradients in the inner penumbrae of MHD simulations of sunspots.

Although, locally, a positive field gradient is clearly visible between the optical depth layers log($\tau$) = $-0.9$, and $0$, this is not the case for the azimuthal averages outside the spot, where the negative gradient contributions dominate the positive ones. Only in the upper node is the average field strength greater than in the deepest layer (see upper panel of Fig. \ref{fld_incl_rad}). 

The strongest negative field gradients concentrated locally near the outer spot boundary (see Fig. \ref{gradb}) are  associated with strong downflows, having the opposite polarity to the umbra. The very strong field strength in these regions in the deepest magnetic layers (some of them reaching up to 4 kG, the value set a priori as an upper limit in the inversion) can probably explain the strong negative field gradients seen in these locations \citep[see, for example,][for details]{van13}.

\subsubsection{Field azimuth and twist in the sunspot}
\begin{figure}[h]
 \centering
 \includegraphics[width=\columnwidth]{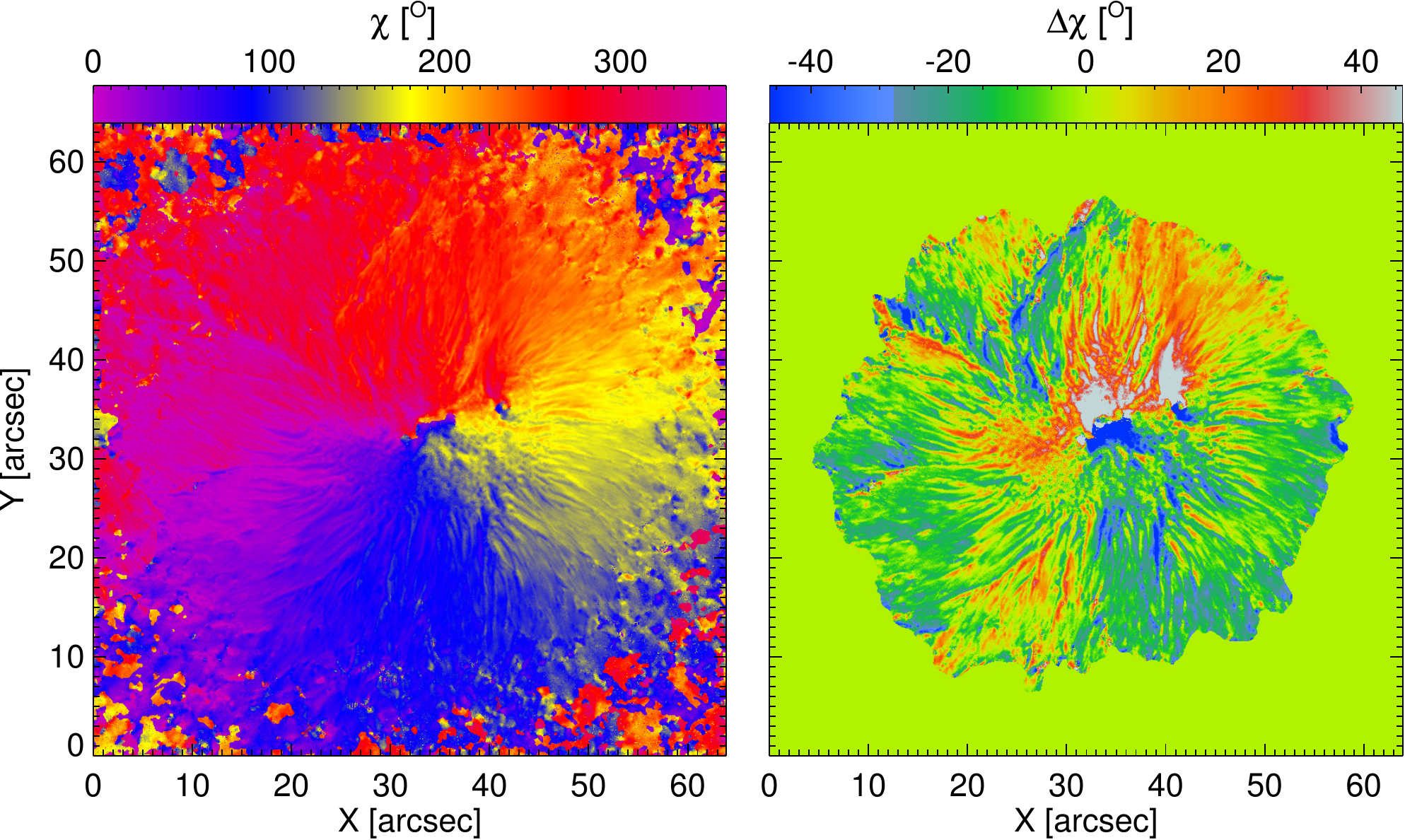}
 \caption{Left panel: field azimuth map at log($\tau$) = $-$0.9. Right panel: twist map calculated from the difference between the measured field azimuth and
   radial directions from the sunspot centre. To avoid contributions of the azimuth from noisy signals in the quiet surroundings, only pixels lying within the sunspot penumbral boundary have been used to compute the twist.}
 \label{azim_maps}
\end{figure}

A field azimuth map, obtained at the middle node (log $\tau$ = $-0.9$), is shown in the left-hand panel of Fig.~\ref{azim_maps}. It displays a generally, radially directed field with a considerable amount of small-scale structure.

\begin{figure}[h]
 \centering
 \includegraphics[width=\columnwidth]{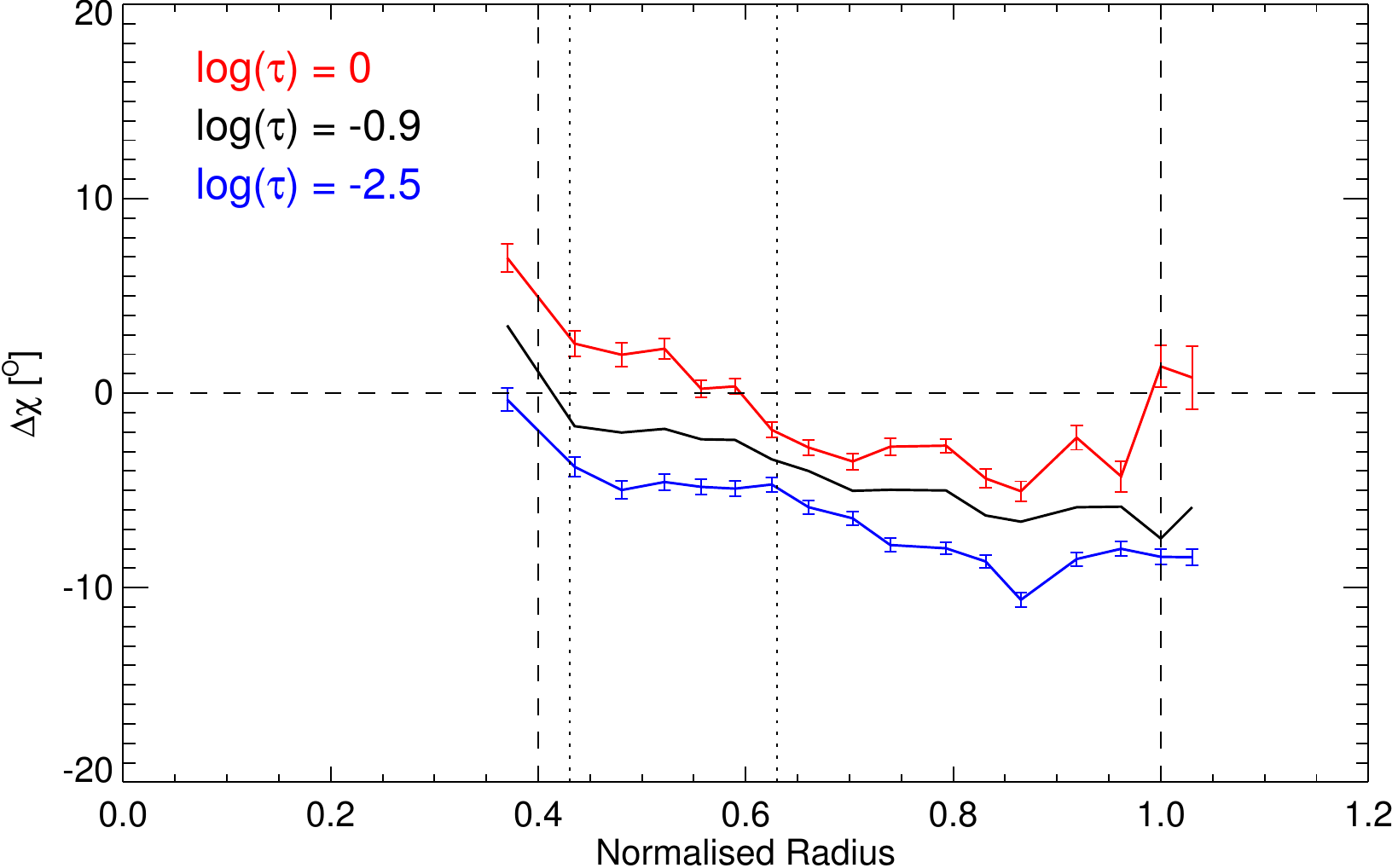}
 \caption{Radial distribution of twist, the azimuthally averaged difference $\Delta \chi$, for the sunspot penumbra at three nodes. All vertical lines are the same as in  Fig. \ref{fld_incl_rad}. The RMS variations are very large along the azimuthal path and not shown here. The error bars represent the standard error of the mean along an azimuthal contour.}
 \label{azimdd_rad}
\end{figure}

To estimate the magnetic twist (average deviation of the field azimuth from the radial direction), we have used only pixels lying within the sunspot boundary. Twist is plotted in the right-hand panel of Fig.~\ref{azim_maps}. We  excluded the quiet-Sun region to avoid the influence of noise and fluctuations. The average twist value of the full sunspot in the bottom layer comes out at $-3.8^\circ$. This value for twist agrees with \cite{tiw09b}, who estimated a twist of the same spot a day before by using a potential field as reference, instead of the radial field. Also, in agreement with these findings, it is clear from the right-hand panel of Fig. \ref{azim_maps} that  much larger and oppositely directed twists dominate locally. Sometimes the local twist reaches a value of $|\Delta \chi | > 90 ^\circ$. The estimated azimuthally averaged twist with radius at the three node positions are plotted in Fig.~\ref{azimdd_rad}. However, the large twist in the umbra seen in Fig. \ref{azim_maps} is partly due to the geometric and magnetic centres of the spot  not being completely co-located. For this reason only, the twist of the penumbral field is plotted in Fig. \ref{azimdd_rad}. 

In the penumbra, there is a trend toward increasing twist with both radius and height. In the middle penumbra, the average unsigned twist in the lower two nodes is generally less than $5^\circ$. Although the variations in the field azimuth along the contours are large, the estimated error in the mean value is not, owing to the large number of pixels along the contours. The error in the mean is indicated in Fig.~\ref{azimdd_rad} and is clearly much smaller than the recovered twist values. However, the error bars are not important for the contours if they have lots of pixels with $|\Delta \chi| > 90^\circ$.

It is also worth mentioning that the field azimuth is the most fluctuating parameter obtained from the inversion. It is also the one most strongly affected by noise. Nonetheless, Fig. 9  does indicate that there could be a gradient in the twist of the sunspot field, both with radial direction and with height.

\subsection{Mass flux}
For a reliable calculation of the mass flux, we require that the density and the LOS velocity be on a geometrical height scale. Not only can the relative geometric height differences between different points in the field of view not be inferred by the inversion code, the density and vertical height scale is also calculated by assuming hydrostatic equilibrium, an assumption that is often not valid in a highly magnetised structure.

When bearing the above limitations in mind, and assuming that $\tau$ unity lies at the same z at all positions in penumbra, a crude estimate of the mass flux over the full sunspot gives about 2.5 times more downflowing mass than upflowing, which agrees with the estimates of \cite{west97,west01a} for a sunspot and with the results of \cite{tiw13} for an averaged sunspot penumbral filament. The amount of excess downflows decreases with height, but even at $\log(\tau) = -2.5$, the excess is a factor of about 1.3.

Although a partial contribution to the downflowing mass from the inverse Evershed flow effect through spines cannot be ruled out, this result suggests that the assumptions made when deducing the mass flux, e.g. hydrostatic equilibrium, log($\tau$) = 0 located at the same z for all penumbral pixels, cannot be true. Therefore, a true geometrical height computation is necessary for  estimating the mass flux over a sunspot reliably. 

Finally, the net mass flux depends on the zero level of the velocity. We  set the velocity at the lowest node in the dark parts of the umbra to zero, since the density is highest at that layer. However, if we force the velocity to be zero at the central node ($\log\tau=-0.9$), then the mass flux excess at $\log(\tau)=0$ reduces to 1.7 (down from a factor of 2.5).

\section{Mutual dependence of physical parameters and sunspot substructures}
In this section, we plot two dimensional (2D) histograms of one physical parameter relative to another and investigate their mutual dependences. Then, by isolating different populations in the histograms, we identify specific substructures of the sunspot.    

\subsection{Field strength $B$ versus inclination $\gamma$}
 \begin{figure*}[htp]
 \centering
 \includegraphics[width=\textwidth]{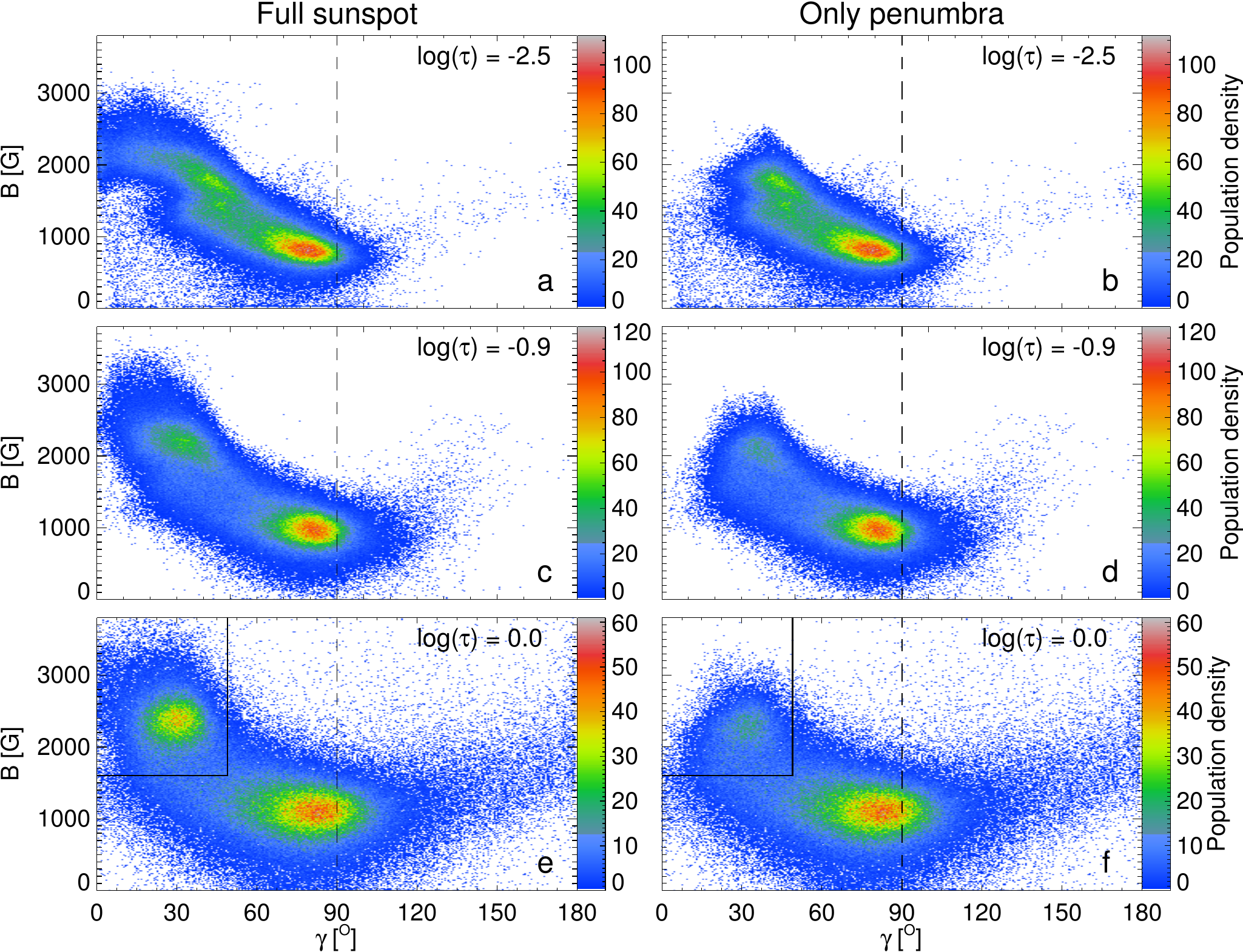}
 \caption{Left panels: 2D scatter plots of $B$ vs $\gamma$ for all points within the sunspot's boundary at the three optical depth nodes (from top to bottom, $\log(\tau)=-2.5, -0.9, 0$). Right panels: the same but only for pixels in the penumbra. The vertical dashed line in each panel represents the location of $\gamma = 90^\circ$. A black box is drawn at log($\tau$) = 0 to identify/outline the umbral pixels, spine pixels in (f) (see Fig. \ref{umbra+spines}(a) and (b) for where these pixels are located in the sunspot).}
 \label{scatter_incl_b}
\end{figure*}

The 2D histograms of $B$ versus $\gamma$ of the full sunspot at the three height node positions are shown in the left-hand panels of Fig.~\ref{scatter_incl_b}. A general anticorrelation between $B$ and $\gamma$ is found at all depths for $\gamma$ $\leq$ 90$^\circ$, in general agreement with \cite{stan97,west01} and \cite{shibu04}. A weak positive correlation, not previously reported, is noticeable at all heights in the range $\gamma$ = 90$^\circ$ - 180$^\circ,$ in spite of the relative scarcity of points.

\begin{figure*}[ht]
 \centering
 \includegraphics[width=\textwidth]{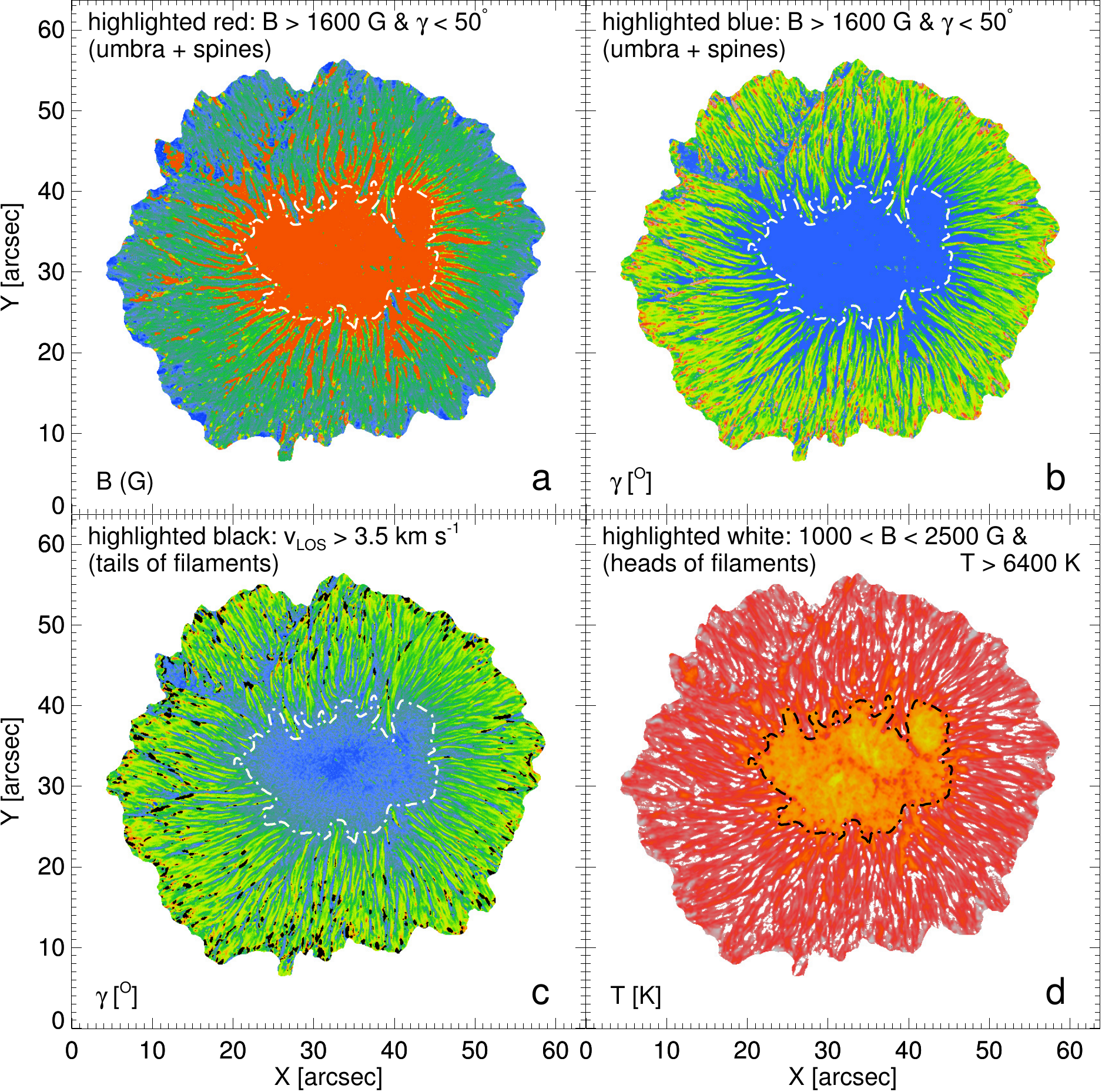}
 \caption{(a) Map of magnetic field strength at log($\tau$) = 0. The pixels with $B >$ 1600 G \& $\gamma<50^\circ$ are highlighted in red (i.e. the pixels in the box outlined in black in Figs. \ref{scatter_incl_b}(e) and 10(f)). (b) Field inclination map at log($\tau$) = 0. with pixels satisfying the same conditions as applied to the field strength map in panel (a) highlighted in blue. The colour scheme for the field inclination map is chosen for consistency with the field inclination map depicted in Fig. \ref{full_maps}. (c) Same map as displayed in (b) but pixels with $v_{\rm LOS} >$ 3.5 \kms\ are highlighted in black, revealing tails of penumbral filaments. (d) Map of temperature at log($\tau$) = 0. Highlighted in white are pixels satisfying $T >$ 6400 K \& 1000 $< B <$ 2500 G, basically representing heads and warm parts of bulks of penumbral filaments. The dash-dotted  contour in each panel represents the umbra-penumbra boundary.}
 \label{umbra+spines}
\end{figure*}

The two peaks in the histograms at all the three heights may be interpreted as belonging to the sunspot umbra (the population in the black box shown for log$(\tau)$ = 0 in Fig.~\ref{scatter_incl_b}e) and the penumbra (the population near $\gamma$ = 90$^\circ$). In an attempt to clarify this picture, we recreated the histograms but only for penumbral pixels, shown in the right-hand panels of Fig.~\ref{scatter_incl_b}. From the residual population of the black box, it is clear that the population, which we were relating to umbra, consists not only of umbra, but also, to a significant extent,  of penumbral pixels. By marking all pixels contained by the black box in Fig.~\ref{scatter_incl_b}(e), in red in Fig.~\ref{umbra+spines}(a), and in blue in Fig.~\ref{umbra+spines}(b), it becomes clear that some of the pixels in the black box belong to the spines of the penumbra, which appear to be true extensions of the umbra into the penumbra. The two populations of umbral/spine pixels and penumbral filament pixels, which are quite distinct at log($\tau$) = 0, increasingly merge with height, although the umbral/spine pixels exhibit much less scatter at log($\tau$) = $-$2.5 than the filament pixels.    

The remaining penumbral pixels, including the second population characterised by nearly horizontal fields with an average field strength of 1 kG, consist of penumbral filaments. Penumbral filaments, whose bulk contains a horizontal field with a strength of $\sim$1 kG \citep{tiw13}, clearly form the dominant part of the penumbra. The right halves of the frames, with $\gamma > 90^\circ$, are populated by the filament tails where the field is stronger, with a polarity opposite to that of the umbra. The opposite polarity is most prominent at log($\tau$) = 0 but is also visible in the higher nodes (albeit for fewer pixels), consistent with the field bending back down into the photosphere.

\subsection{Temperature $T$ versus magnetic field strength $B$}

 \cite{bier41} was the first to realise that the presence of a strong magnetic field in a sunspot could be responsible for its darkness. \cite{alfv43} predicted a relationship between the magnetic field strength and temperature, which has been extensively investigated both theoretically by, for example, \cite{chit63,dick70,cowl76,malt77} and \cite{spru90}, and observationally by, for example \cite{chou87,pill90,kopp92,pill93,west01} and \cite{shibu04}. However, the relationship between $B$ and $T$ of sunspots, particularly the correspondence between different parts of scatter plots found by these authors and features of sunspots, has not yet been understood \citep[see][for a detailed review of the subject]{sola03}.

\begin{figure*}[htp]
 \centering
 \includegraphics[width=\linewidth]{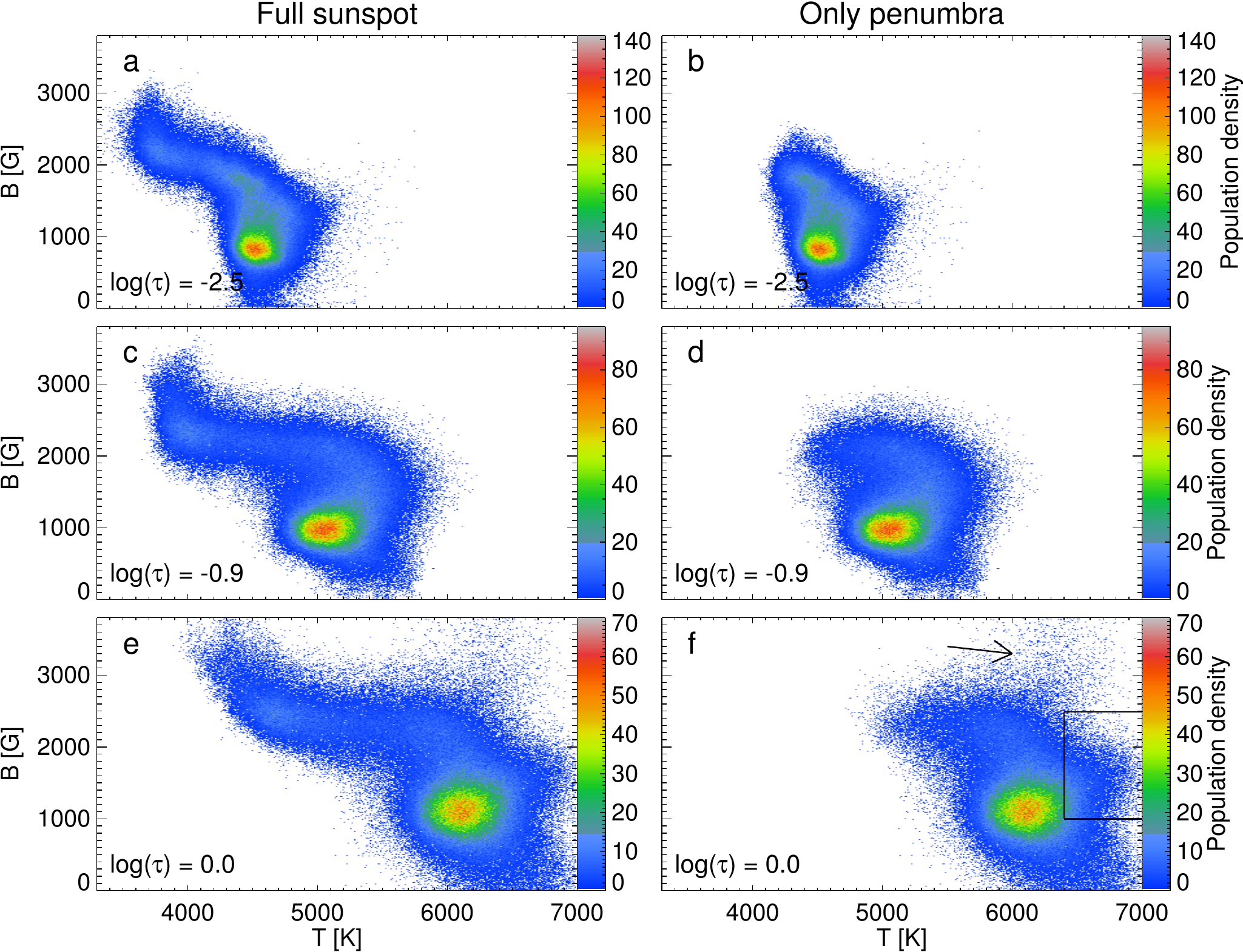}
 \caption{Left panels: 2D scatter plots of $B$ vs $T$ for all pixels. Right panels: The same but only for penumbral pixels inside the sunspot, shown at the optical depths of the three nodes. The black box in (f) outlines the heads of some of penumbral filaments (see, Fig. \ref{umbra+spines}(d) to find the spatial locations of these pixels). The arrow in (f) points to the population that belongs to tails of penumbral filaments that are subsets of highlighted pixels in Fig. \ref{umbra+spines}(c)).}
 \label{scatter_T_B}
\end{figure*}

The 2D histograms of $T$ versus $B$ in Fig.~\ref{scatter_T_B} allow us to revisit the thermal-magnetic relationship of sunspots. The temperature and its spread over different parts of the sunspot decrease rapidly with height. At first glance, the plots for the full sunspot in the left-hand panels appear nearly identical to the results obtained by \cite{kopp92,sola93a,pill93,stan97,west01,penn03a} and \cite{shibu04}. However, a closer look reveals noticeable differences, mainly in the right-hand parts of the plots, which are expected to be populated by penumbral pixels. To confirm this identification, we plotted the same histograms at the three heights but only for penumbral pixels, shown in the right-hand panels of Fig.~\ref{scatter_T_B}. The most densely populated area is at around 1 kG field strength and a temperature of 5~900 - 6~400 K at log($\tau$) = 0.

Although the general impression of Fig.~\ref{scatter_T_B} is of an anti-correlation (also in the penumbra, where the correlation coefficient is around $-$0.4), we do find a localised weak positive correlation between the two parameters, particularly visible in the region outlined in Fig. \ref{scatter_T_B}(f). The positive correlation between the temperature and the magnetic field strength is probably caused by the brighter heads of the penumbral filaments, which are brighter than their surroundings and often contain a stronger magnetic field \citep{tiw13}. To confirm this, we identified the locations of the pixels lying in the black box, and highlighted them in Fig.~\ref{umbra+spines}(d), which confirms that they are, to a significant extent, the locations of the heads of penumbral filaments; some pixels, e.g. from the bulk of filaments, also appear to be present. An additional scatter plot (not shown here) of $B$ vs $T,$ showing only highlighted pixels, confirms this weak positive correlation. However,  not all the heads of  penumbral filaments are highlighted by the selection criteria described in Fig. \ref{umbra+spines}(d). 
For example, some heads may contain a temperature of 6400 K or less, which  then fall outside the box towards lower temperatures and are not highlighted in Fig. \ref{umbra+spines}(d).  

Another anomalous, rather scattered, population in the Fig.~\ref{scatter_T_B}(f) is formed by the pixels above 2500 G and 5500 K (indicated by an arrow). This is formed by the tails of penumbral filaments, which have stronger fields and are somewhat darker than the heads of filaments. As shown by \cite{tiw13}, the tails of filaments show an enhancement in temperature,   sometimes becoming hotter than parts of their bulk \citep[see Fig. 5 of][]{tiw13}. The panels (e) and (f) of Fig. \ref{scatter_T_B} show that the strongest fields in the sunspot  belong either to the umbra or to the tails of penumbral filaments. 

The highlighted pixels for tails of penumbral filaments in Fig. \ref{umbra+spines}(c), by the condition $v_{\rm LOS} > 3.5$ \kms, as described later, also accommodate the pixels with $T (> 5500 K)$ and $B (> 2500 G)$ of the population indicated by an arrow in Fig. \ref{scatter_T_B}(f).     

\begin{figure*}[ht]
 \centering
 \includegraphics[width=\linewidth]{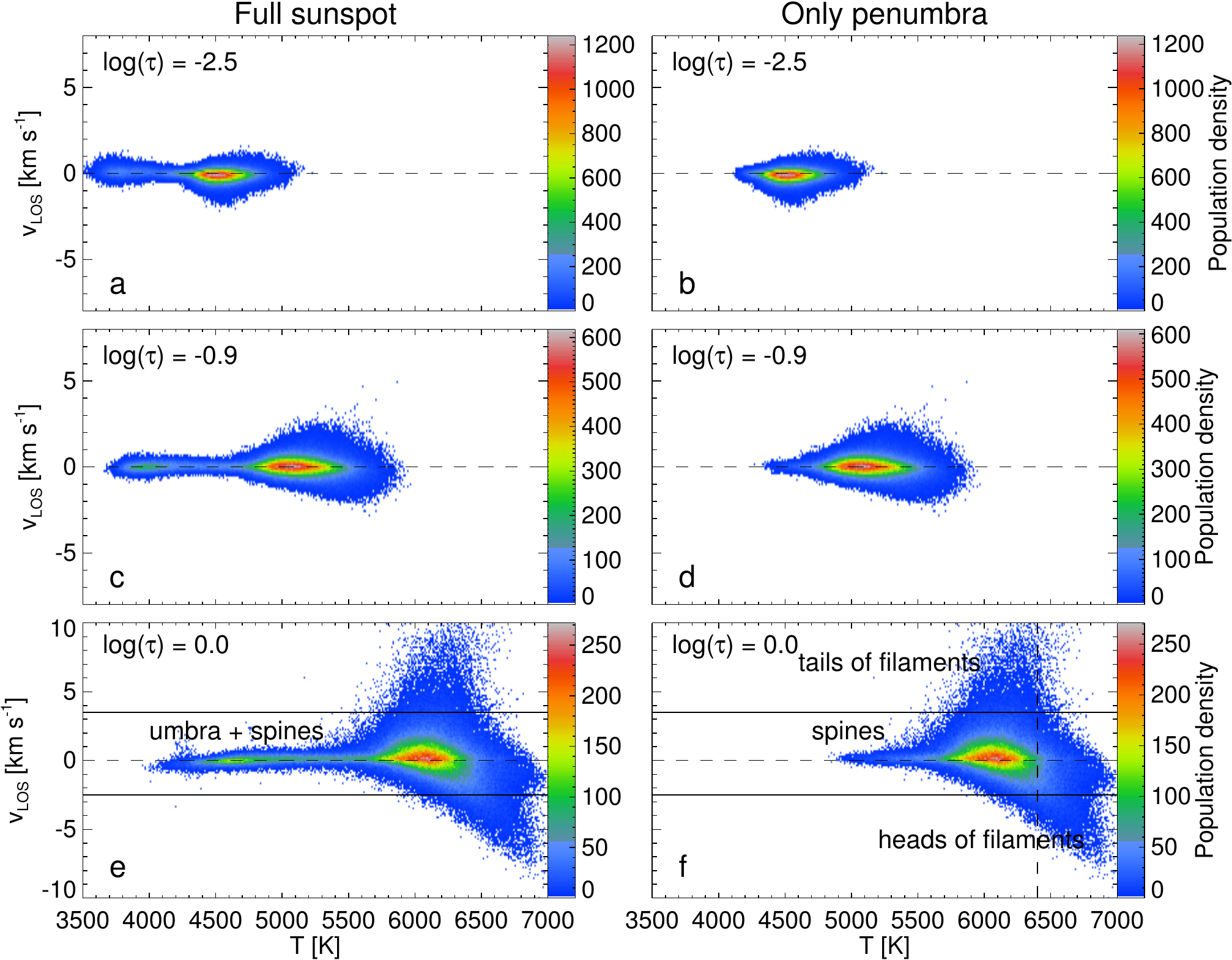}
 \caption{Left panels: 2D scatter plots of $T$ vs $v_{\rm LOS}$ for all points within the sunspot boundary at the three optical depth nodes. The horizontal dashed line in each panel indicates the zero velocity level. In each panel, the points at lower temperature belong to the sunspot umbra and also to spines, while those at higher temperature belong only to the penumbra. Right panels: the same, but only showing penumbral pixels. The two horizontal solid lines are used to separate, in an approximative manner, the heads (lying below the lower solid line) and tails of penumbral filaments (above the upper solid line) from the other points. The dashed vertical line in (f) represents $T$ = 6400 K, a condition applied to isolate the heads of penumbral filaments in Fig. \ref{umbra+spines}(d), although it also includes points from the bulks of the filaments (points to the right of the vertical dashed line lying between the two horizontal solid lines).}
 \label{scatter_T_velos}
\end{figure*}

\subsection{LOS velocity $v_{\rm LOS}$ versus temperature $T$}

The upflowing gas is expected to be hotter than the downflowing gas if these flows are driven by convection.
Two dimensional histograms of $v_{\rm LOS}$ versus $T$ for the full sunspot at the three node positions are displayed in the left-hand panels of Fig. \ref{scatter_T_velos}. The majority of the coolest pixels (mostly belonging to the umbra) are concentrated around zero velocity at all  three heights (the horizontal branch of the histograms to the left), with some deviations in the deepest node because of umbral dots \citep[e.g.][]{riet13} and faint light bridges \citep[e.g.][]{lagg14} containing up- and downflows.

The broad distribution of points above roughly 5500 K in all panels is the penumbral contribution, as confirmed by similar histograms of only the penumbral pixels and as depicted in the right-hand panels of Fig.~\ref{scatter_T_velos}. It is clearly visible  that the upflows at $\log(\tau)=0$ are, on average, hotter than the downflows. The average temperatures of up- and downflows are $\sim$ 6300 and 6050 K, respectively, for full penumbral pixels, and 6600 and 6200 K for the pixels outside the two solid horizontal lines. There is a weak tendency for the strongest downflows to be warmer than the weak downflows. 

The hot upflows and cool downflows, particularly the extended populations below and above the horizontal lines, respectively, in the panels (e) and (f) of Fig. \ref{scatter_T_velos} can readily be associated with the heads and tails of penumbral filaments. To confirm that the extended population above the upper horizontal line in Fig. \ref{scatter_T_velos}(f) belongs to the tails of penumbral filaments, we highlighted those pixels in Fig.~\ref{umbra+spines}(c). Similarly, we tested if the population below the lower horizontal line in Fig. \ref{scatter_T_velos}(f)  belongs to heads of penumbral filaments, and found that they are accommodated in the highlighted pixels in Fig.~\ref{umbra+spines}(d). The pixels below the lower solid line in Fig. \ref{scatter_T_velos}(f) are only a small subset of the pixels highlighted in Fig. \ref{umbra+spines}(d). 

\begin{figure*}[ht]
 \centering
 \includegraphics[width=\linewidth]{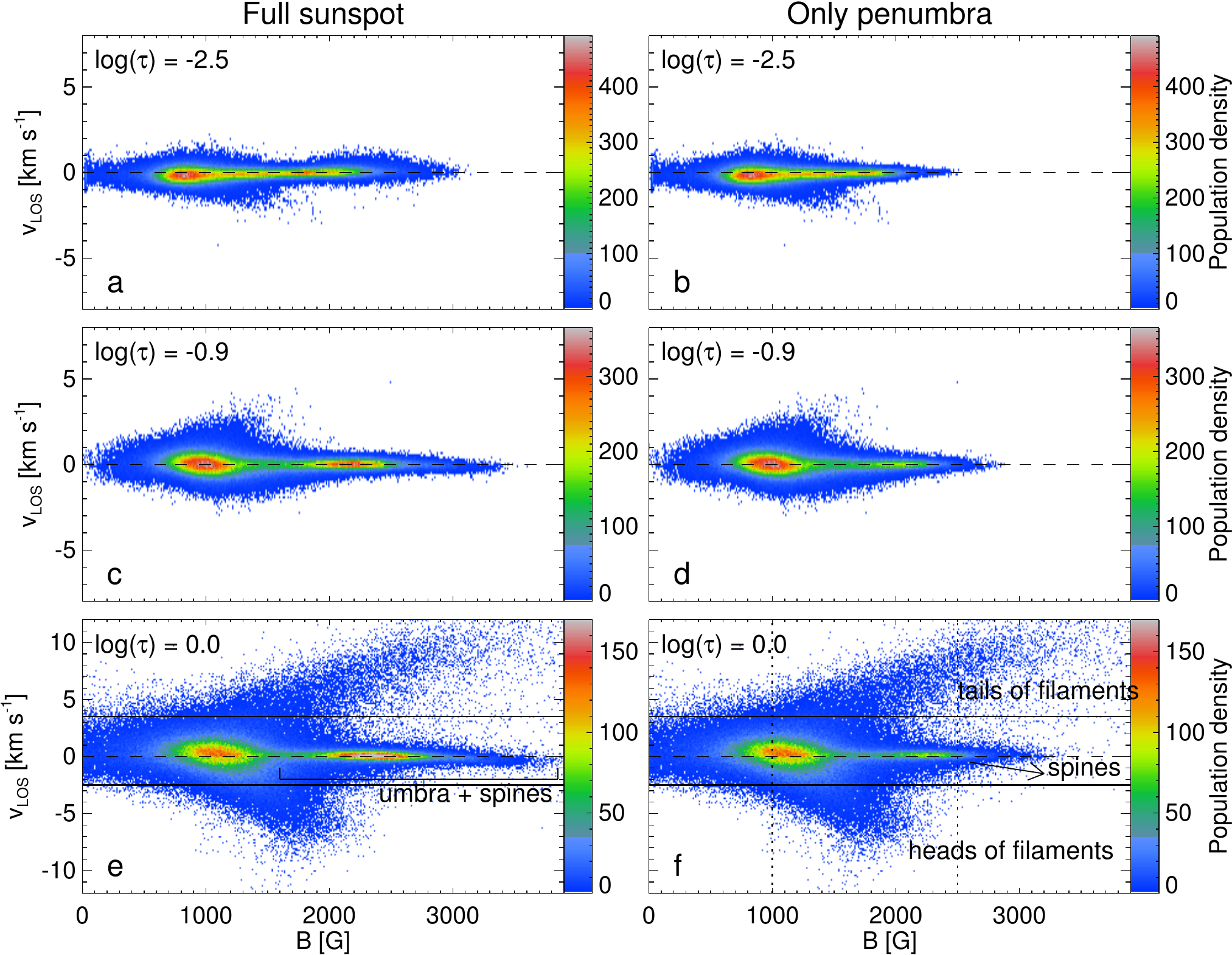}
 \caption{Left panels: 2D scatter plots of $B$ vs $v_{\rm LOS}$ for all points within the sunspot boundary at the three optical depth nodes. The horizontal dashed line in each panel indicates the zero velocity level. Right panels: the same, but only showing pixels from the penumbra of the sunspot. The two horizontal solid lines in  panels (e) and (f) indicate thresholds of the velocity used to identify, in an approximative  manner, the heads and tails of the penumbral filaments. The two vertical dashed lines in panel (f) bound the field strength range $1000 < B < 2500$ G, a condition used to identify heads of penumbral filaments in Fig. \ref{umbra+spines}(d).}
 \label{scatter_B_velos}
\end{figure*}

In Figure \ref{scatter_T_velos}(f), the spines are cooler and contain very low velocities. These spines get cooler with height and other penumbral pixels merge with them. In the topmost layer, in Figure \ref{scatter_T_velos}(b), unexplained weak upflows can be seen that appear to be cooler than the downflows. Such a phenomenon is well-known in the quiet Sun, and known as  reversed granulation, with granules becoming cooler and intergranular lanes becoming hotter above log optical depth of roughly $-$2. We speculate that this (weak) reversal, seen at $\log(\tau)=-2.5$ in the penumbra, is the signature of the same effect (described by, for example, \cite{nord09} and is the result of the adiabatic cooling of the expanding upflowing gas).

\subsection{LOS velocity $v_{\rm LOS}$ versus magnetic field strength $B$}

Figure~\ref{scatter_B_velos} shows the 2D histograms of $v_{\rm LOS}$ vs $B$ for the whole spot (left-hand panels), and for the penumbra only (right-hand panels). The most prominent features in the histograms are the two branches towards the upper and lower right corners of the domain, most clearly visible at $\log(\tau)=0$. These are readily associated with the tails and heads of penumbral filaments, respectively, and suggest that the strongest magnetic fields are not only found in the umbra, but also in the deepest layers of the fastest downflows, which are expected to lie in the tails of penumbral filaments, based on the work of \cite{van13} and \cite{tiw13}. 

To confirm the above prediction of the positions of pixels belonging to the heads and tails of penumbral filaments, we   overplotted separately all the pixels outside the two solid horizontal lines shown in Fig. \ref{scatter_B_velos}(f). As shown in Fig. \ref{umbra+spines}(c), the pixels with $v_{\rm LOS} >$ 3.5 \kms\ create a good match with the tails of filaments. The pixels with $v_{\rm LOS} < -2.5$ \kms\ are completely accommodated within the highlighted pixels of Fig. \ref{umbra+spines}(d).   

There is a significant difference of 300 G (1~500 G vs 1~800 G) in the field strengths of average upflows ($-$3.9 \kms) and downflows (5.8 \kms) computed outside the solid lines, the latter being stronger. This difference can generate an outflow via a siphon flow, if it refers to the same geometrical height \citep{meye68,mont97}. However, the higher average temperature (6600 K) of upflows than downflows (i.e. 6200 K) supports the convection mechanism as the driver of the Evershed flow. Thus, a conclusive statement about the driving mechanism of the Evershed flow cannot be made. 

The reduced population of spine pixels in the left- and right-hand panels of Fig. \ref{scatter_B_velos} at all heights again shows that the sunspot umbra and spines exhibit similar physical properties. A rapid reduction in both the up- and downflows with increasing height is in agreement with the literature \cite[see for example,][and references therein]{west01a,sola03}. A weak tendency of a shift from overall dominant downflows to weak dominant upflows with increasing height can be observed, however.

\section{Discussion}

We have performed depth-dependent, spatially coupled inversions of a disk-centered sunspot, observed by Hinode SOT/SP, and have presented the depth-stratified thermal, velocity, and magnetic atmospheric structure of it. We have also looked at how the small-scale structures fit in with the global behaviour of the sunspot and have introduced simple thresholds in individual parameters to isolate some of the fundamental constituents of the sunspot penumbrae. In the following, we interpret our results and discuss them in the context of earlier work available in the literature.

\subsection{Global properties}\label{dis_gp}
The continuum intensity and temperature at $log(\tau)$ = 0 display a very similar trend, an obvious increase with radius starting from the centre of the spot. The horizontal gradient gets smoother for higher layers. Results imply that the vertical temperature gradient in the sunspot penumbra is slightly smaller than that in the quiet Sun, but is much larger than that in the umbra. These results are in good agreement with findings in the literature \cite[see, for example,][]{schr71,malt86,coll87,lites93,stan97,west01,sola03,trit04}. 

The azimuthally averaged magnetic field strength decreases with radial distance from the centre of the sunspot (2800 G) to the outer penumbral boundary (700 G at $\tau$ unity level), in qualitative agreement with earlier findings, e.g. \cite{west01,shibu03,borr11}. However, some quantitative differences can be seen, beyond the well-known difference in the maximum $B$ of spots \citep[\eg][]{schad13}, which is known to depend on the size of spots. For example, the averaged field strength values of about 2300 G and 2500 G for umbra, and 500 G and 700 G for outer penumbral boundaries, found by \cite{west01}, and \cite{shibu03}, respectively, might be due to the variation of properties from one sunspot to another (see also \cite{sola03}) but may also reflect differences in the spectral lines used and the spatial resolution of the data. The maximum $B$ in our sunspot is comparable to that of the sunspots with similar areas in the work of \cite{schad13}.   

The vertical gradient in the photospheric layers of the umbra of $-$1.4 G km$^{-1}$ is comparable to that obtained in the most recent MHD simulations (e.g. $-$1.5 G km$^{-1}$: M. Rempel, 2013, private communication). In the literature, vertical field gradient values in sunspot umbrae varying from $-$4 G km$^{-1}$ \citep{west98} to $-$1.5 G km$^{-1}$ \citep{coll94,west01,balt08} have been reported, but see also \cite{schr71} for smaller vertical field gradients of about $-$0.5 G km$^{-1}$. For a detailed review of earlier results, see \cite{sola03}. 

The positive vertical magnetic field gradient observed in the middle part of the sunspot penumbra is partly caused by the partial cancellation of the Stokes $V$ signal at the unresolved interface between the spines and the partly, oppositely directed field, found at the edges of the filaments, near their heads. However, this is also an artefact of the highly corrugated iso-$\tau$ surface in the inner penumbra, see \cite{balt08,balt13,josh14,josh15}. The last two investigations present a particularly thorough analysis of such observations.

The main indication of a magnetic canopy in our inversion results is the presence of a stronger field at the upper node than at the two lower nodes. Such a signature is only found outside the sunspot. This agrees with earlier findings of \cite{giov80,giov82,sola92,sola94,sola99,adam93,balt08}, but partly differs from the results of \cite{west01}, who found a canopy-like structure starting from the middle of the penumbra and continuing outside the penumbral boundary of their spot. A similar canopy-like structure to that found by \cite{west01} was reported more recently by \cite{borr11}. The fact that \cite{shibu03} found no canopy structure, even outside the visible boundary of their sunspot, is probably due to the much lower formation height of the 1.56 $\mu$m lines that they used and, as such, need not be inconsistent with our results. The sunspot canopies inferred from the 1.56 $\mu$m lines by \cite{sola92,sola94} were deduced using another technique, and not from the explicit height-dependent inversion of spectral lines. Although not so clearly marked, a canopy structure outside the sunspot boundary, similar to that obtained here, was seen by \cite{balt08}.

The larger azimuthally averaged inclination of the field by about 10$^\circ$ (i.e. on average more horizontal field) in the lower layers throughout the spot is consistent with the findings of \cite{west01} and \cite{borr11}, who attributed this to the field canopy structure starting in the inner penumbra. However, it disagrees with the finding by \cite{shibu03} in which the field becomes more horizontal with height. This disagreement could be caused by the different formation height of the IR lines used by \cite{shibu03} or by the stronger influence of straylight on their results.

Although we expect the field to be more vertical with height in a simple monolithic flux-tube model of a sunspot, a change of 10$^{\circ}$ within the lower photosphere is too large to be explained in the context of a homogeneous flux-tube model. Instead, this can be understood in terms of small-scale structure. Umbral dots and light bridges that contain stronger horizontal fields than in the umbral background, are visible only in the deepest layers \citep[see for example,][for umbral dots and light bridges, respectively, and references therein]{riet13,lagg14}, but do not contribute to the averaged inclination at higher layers. Similarly, in the penumbra, the restriction of the mainly horizontal-field penumbral filaments to the low photosphere and the wrapping of the magnetic field around them \citep{borr08,tiw13} leaves only a relatively vertical field in the upper layers of the solar photosphere, as proposed by \cite{sola93}.

The general inverse relationship between the magnetic field strength, $B$, and its inclination ,$\gamma$, is well known from earlier observations with lower spatial resolution \citep{sola93a,stan97,west01,shibu04}. We revisit this relationship with our high resolution data set and find a stronger anti-correlation between these two parameters than has been obtained by earlier researchers (e.g.\ \cite{west01}). We also find the presence of two populations, clearly separated at log($\tau$) = 0: one of umbral and of penumbral spine pixels, the other of penumbral filament pixels. At higher layers the two populations merge, possibly because the spines expand and their field wraps around the filaments.    

Another novel feature that we  found is a positive correlation between the field strength and the inclination for $\gamma > 90^\circ$, most particularly at log($\tau$) = 0. We attribute this to the tails of the penumbral filaments, which contain opposite polarity and a stronger magnetic field \citep{tiw13}. 

It has been known for decades that the darker umbral regions contain more intense magnetic fields \citep{kopp92,pill93,stan97,west01,sola03}. We confirm the general relationship at all heights.  
In qualitative agreement with  earlier researchers, we find that the darker, and therefore cooler, regions contain stronger fields in the umbra at all heights, with a dependence roughly in the shape of a raised elephant's trunk (see Fig. \ref{scatter_T_B}). This is particularly visible in the upper two node positions, indicating that the temperature is nearly constant for a range of the strongest umbral fields.

In the penumbra, however, this relation is not universally valid. The heads of penumbral filaments contain strong fields \citep{tiw13}, yet they are bright, giving rise to a positive correlation between the temperature and magnetic field in the sunspot penumbra. In the early literature, this positive correlation between temperature and field was mis-interpreted as an indication that the spines were the brighter parts of the sunspot penumbra (e.g. \cite{west01}, see a review by \cite{sola03}, and \cite{tiw13} for detailed clarification).

The average LOS velocity shows dominant upflows (peak at $\sim -300$ \ms) in the inner penumbra and downflows (peak at $\sim 1300$ \ms) in the outer penumbra, in qualitative agreement with earlier observations \citep[e.g.][]{rimm95a,schl00,west01a,trit04,sanc07,ichi07,fran09}. This observation supports the scenario of the Evershed flow rising in the inner penumbra and sinking in the outer penumbra. This average flow pattern is the sum of the flows in individual penumbral filaments. The upflows are concentrated in the heads, the downflows in the tails of individual filaments \citep{tiw13}. Since the heads are closer to the umbra and the tails closer to the outer boundary of the spot, azimuthal averaging produces upflows in the inner and downflows in the outer penumbra. The upflows along the axis of the filaments and downflows along their sides \citep{tiw13} partly average out during the azimuthal averaging and contribute less to the global radial trend. 

An average downflow of about 350 \ms\ is seen over the sunspot penumbra at $\log(\tau)=0$, caused by the presence of strong downflows at the outer penumbral boundary. The strong downflows are found to continue outside the outer penumbral boundary, even reaching a maximum there in the azimuthal averages (see Fig. \ref{vel_temp_rad}). This is in agreement with the results of, for example, \cite{shee72,dere90,boer92} and \cite{sola94}, that a part of the mass transported by the Evershed flow continues beyond the sunspot's visible boundary.

We can see in Fig. \ref{vel_temp_rad} that the azimuthally averaged velocity in the two top nodes is not zero in the umbra, but instead shows an average downflow that increases with height. Since the density decreases rapidly with height, any downward mass flow would result in such a height dependence of the velocity. This downflow could be a result of the inverse Evershed flow \citep{malt75}. 

The clear pattern of hotter upflows than downflows in the sunspot penumbra (by an average difference of about 400 K between heads and tails of penumbral filaments) is consistent with the transport of heat by convective motions in the sunspot penumbra, as has variously been proposed to explain the observed penumbral brightness \citep{dani61b,chit63,meye74,weis91,jahn94,schu01,weis02,weis04,thom08,remp09b}. 
At the same time, however, the downflow speed is correlated with field strength, and strong downflows are associated with stronger fields than upflows (by an average difference of about 300 G between heads and tails of filaments). This result is consistent with the requirements of the siphon flow model \citep{meye68,mont97}. Therefore, the up- and downflows in the penumbra exhibit properties that make them partly consistent with both  main rivalling theories of the Evershed flow (the flux tube: \cite{sola93,schl98a,schl98b,borr05}, and field-free gap models: \cite{spru06,scha06}). One caveat is that, to test the siphon flow, we need to know $v$ and $B$ at a given geometrical height, which is not fulfilled (to an unknown extent) by the present data.   

The average azimuthal twist within the sunspot penumbra for the deepest layers was found to be $-3.8^\circ$, which is in agreement with the twist found by \cite{tiw09b} for this spot (as observed one day earlier), and estimated by computing the spatially averaged signed shear angle, which returns the twist of a spot, irrespective of its shape and force-free nature of the field. The twist in the spot increases with the radius as well as with height. The increase in the twist with height can be understood as a result of the expansion of a twisted flux tube that leads to an increase in the azimuthal component of the magnetic field, in turn resulting in an increase in the measured twist   from $\tan^{-1}(B_{\psi}/B_r)$ according to \cite{venk09}; \citep[see, e.g.][]{parker74,parker75,parker79}. The increase in the twist with radius might be a result of the Coriolis force acting on the outflowing material of the Evershed flow. This effect becomes stronger at larger radii in the penumbra because of the weaker average field \citep[see, for example,][]{pete96}. 
A caveat worth mentioning on the determined twist  is given by the results of \cite{bueh13}, who find systematic, unexplained positive twists in all features of both the magnetic polarities. 

\begin{figure*}[th]
 \centering
 \includegraphics[width=\textwidth]{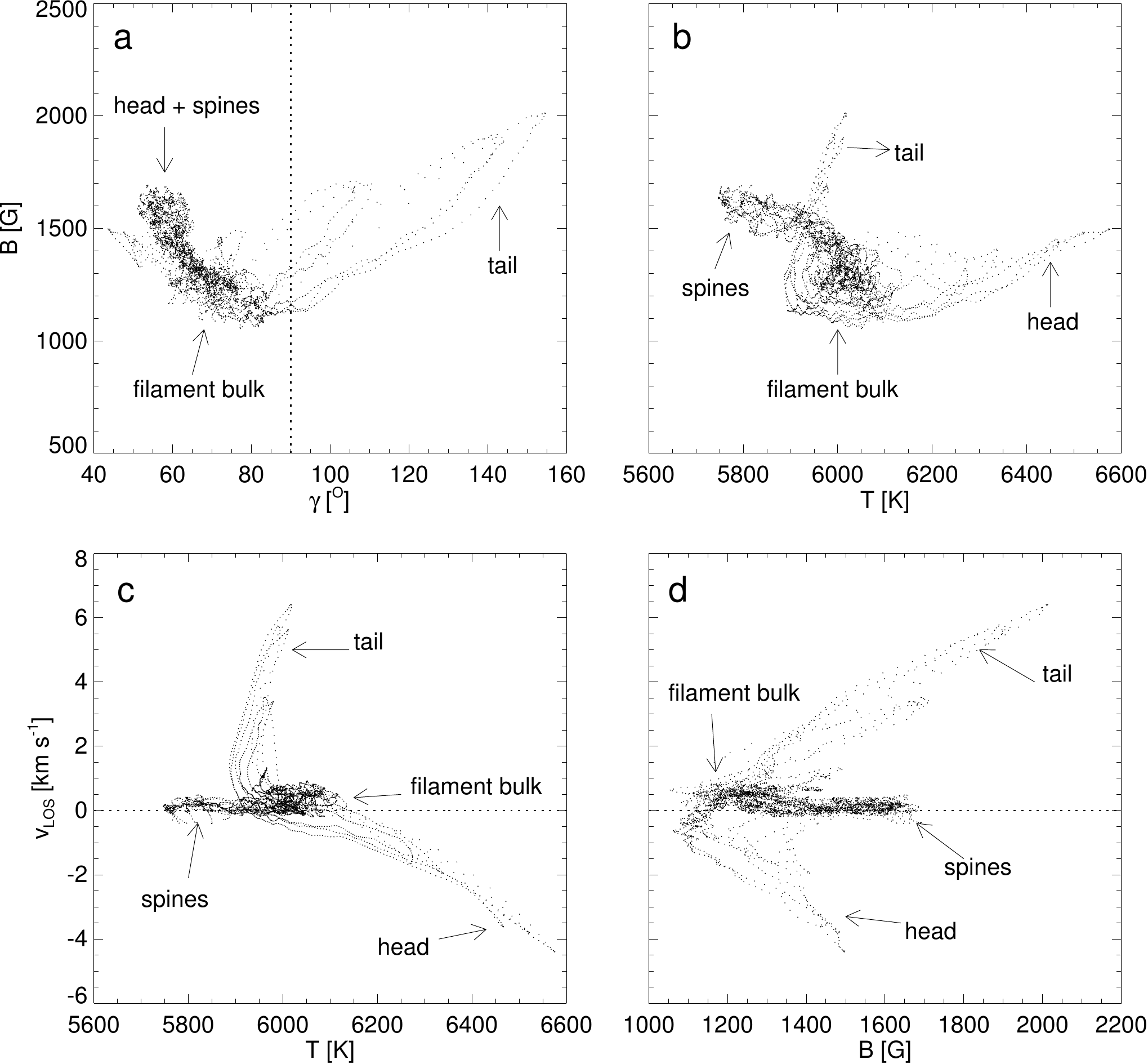}
 \caption{Scatter plots of different physical parameters of the standard penumbral filament studied by \cite{tiw13}. The selected area also contains pixels belonging to the surroundings of the filament (spines). All plots are for the height log$(\tau)$ = 0. The four panels show scatter plots between: (a) the field strength $B$ and inclination $\gamma$, (b) $B$ and temperature $T$, (c) the line-of-sight velocity $v_{\rm LOS}$ and $T$, and (d) $v_{\rm LOS}$ and $B$. The vertical line in (a) indicates $\gamma$ = 90$^\circ$; horizontal lines in panels (c) and (d) represent $v_{\rm LOS} = 0$. Different parts of the filaments are identified in each plot.}
 \label{scatter_fila}
\end{figure*}

A crude estimate of the mass flux over the full sunspot suggests that the downflowing mass is about 2.5 times larger than the upflowing mass. This result is in agreement with that obtained by \cite{west97} for a full sunspot and by \cite{tiw13} for a standard penumbral filament. The latter authors suggest that this excess downflow can be attributed to the corrugation of the surfaces of constant optical depth within penumbral filaments. In contrast, \cite{pusc10} find the upflowing mass flux to be five times larger than the downflowing mass flux in part of a sunspot penumbra. As described in \cite{tiw13}, this inconsistency could be because  \cite{pusc10} evaluate mass flux on the inner part of penumbra, where upflows dominate due to the dominant presence of heads of penumbral filaments, and also partially because they take the diskward part of the penumbra, which is blue-shifted owing to the Evershed flow. Another factor that could contribute to the difference is that \cite{pusc10} estimate the mass flux on a  surface of constant geometrical height, which they deduce based on the requirement of div${\bf B} = 0$, whereas the mass flux was calculated on a surface of constant optical depth by \cite{west97} and by us.

\subsection{Constituents of the penumbra} 

The filamentary nature of the sunspot penumbra has been investigated for more than 50 years (e.g.  \cite{dani64,moor81,moor81r,titl93,lites93,rimm95,lang05,ichi07a,borr08,borr11,josh11,scha11,scha12,scha13,tiw13}). One proposal, given by \cite{lites93}, divides the sunspot penumbra into two components: spines, the more vertical and stronger fields; and interspines: more horizontal and weaker fields. Our results clearly show (see Fig. \ref{umbra+spines}) that the spines have properties similar to the umbra, but with increasing field inclination and decreasing field strength radially outwards. It is worth mentioning that these locations (umbra and spines)  more closely satisfy force-free conditions on the photosphere than other parts of sunspots \citep{tiw12}.

The remainder of the penumbra resembles the standard penumbral filament presented by \cite{tiw13}, who show that the penumbral filaments display similar properties to stretched granules. They find that penumbral filaments contain strong upflows at their heads and downflows at their tails. The upflows continue along the central axis of filaments to more than half of their lengths, surrounded by weak downflows on both the sides of filaments, signatures of which were already reported by, for example, \cite{josh11,scha11}, and \cite{scha12}. The locations of these lateral downflows near heads of filaments were found to contain a field of opposite polarity  to that of umbra, spines, and heads of filaments in one third of the penumbral filaments that \cite{tiw13} studied, in agreement with the results of \cite{remp12,ruiz13} and \cite{scha13}.
Additional features, often counted as constituents of the penumbra, such as penumbral grains, turned out to be heads of penumbral filaments. 

In Fig.~\ref{scatter_fila} we display scatter plots of a number of physical quantities at log($\tau$) = 0 from the standard penumbral filament created by \cite{tiw13}. By choosing a fixed width for all filaments contributing to the standard filament, the latter also includes parts of neighbouring spines at its sides and around  its head. The contributions of different parts of the filament to the clouds of data points are marked by arrows.

A comparison of the scatter plots in Figs. \ref{scatter_fila}(a), (b), (c), and (d) with Figs~\ref{scatter_incl_b}(f), \ref{scatter_T_B}(f), \ref{scatter_T_velos}(f), and \ref{scatter_B_velos}(f), respectively, reveals strong similarities, supporting our speculation that penumbrae can be described as a combination of spines and filaments. The main difference is the lower scatter in Fig. \ref{scatter_fila}, which is due to the averaging carried out to produce the standard filament. The smoothness, e.g. of spines, and the fact that many points describe curves in scatter plots of the standard filament, is a result of averaging and bicubic spline interpolation along the length of filaments \citep[for details, see,][]{tiw13}. In addition, the surroundings of filaments in the outer penumbra contain fewer spines. Instead filaments lie directly next to each other, thus diluting the values for spines and making them smoother in the scatter plots.

\section{Conclusions}
Using the SPINOR inversion code, we performed a spatially coupled, depth-dependent inversion of a sunspot, observed almost at solar disk centre by the Hinode (SOT/SP). We investigated the spot's thermal, velocity and magnetic structure. The average vertical field gradient in the sunspot umbra near the deepest node is $-$1.4 ~G km$^{-1}$, which is in agreement with that found in the recent MHD simulations. The azimuthally averaged magnetic field and inclination show a global general trend as identified in earlier work: the field strength decreases with the radial distance from spot centre and with height, and the field inclination increases with radius, but decreases with height. A positive vertical gradient of the magnetic field strength in the middle of the penumbra is found that could be either the result of a strongly corrugated optical depth surface in the penumbra, the signal cancellation of unresolved fine structure in the deep photosphere, or a combination of both \cite[][]{josh15}. We also find a magnetic canopy structure outside the visible boundary of the penumbra. The temperature displays a slightly stronger vertical gradient in the quiet Sun than in the penumbra, but this is much stronger than in the umbra. The flatter $T(\tau)$ profile of the umbra is in agreement with earlier investigations \cite[see, for example,][and references therein]{sola03}.  

The strongest fields are found in the darkest regions of the umbra, but also in the nearly vertical, opposite polarity flux concentrations in the tails of penumbral filaments. Horizontal fields in the penumbra show an average field strength of 1 kG, in agreement with the strength obtained in the bulk of a standard penumbral filament by \cite{tiw13}.

The well-known general trend for the magnetic field to be strongest in the darkest regions is found in the umbra and in spines. However, in penumbral filaments, in particular at their heads and in their tails, the opposite trend is found. The field azimuth shows an average negative twist of $< 5^\circ$. The twist in the penumbra increases both with radial distance from spot center and with height.

Azimuthally averaged LOS velocities at the deepest layer display an upflow (of 300 m s$^{-1}$) in the inner penumbra, where the number of filament heads harbouring strong upflows is largest. In the outer penumbra, where the tails of filaments outnumber the heads, an average downflow of 1300 m s$^{-1}$ is observed. On average, the sunspot penumbra contains a downflow of $\sim$350 \ms. All over the sunspot penumbra, upflows are found to be hotter (by about 400 K)  than downflows, thus qualitatively supporting the thesis that convection accounts for the observed brightness of the penumbra and is responsible for driving the Evershed flow. However, we also find that strong downflows are, on average, associated with stronger fields (by about 300 G) than upflows, as required by the siphon flow model. Therefore, these data do not in themselves allow us to distinguish between these two mechanisms, partly because the actual geometric height corresponding to a given $\tau$ level is unknown.  

A crude estimate of the mass flux over the full sunspot reveals that the total downflow of mass exceeds the total upflowing mass by a factor of $\sim$2.5. This imbalance is probably a consequence of the optical depth corrugation. Since the upflows are associated with hotter and less strongly magnetized gas than the downflows, we expect that the upflows have been detected at a greater height than the downflows. Thus any upflowing mass that turns over and flows  down again between these two levels is not detected, although the associated downflow is seen, giving rise to an excess in the downflowing mass. 

We find that the spines in the penumbra display qualitatively similar characteristics to those in  the umbra. They are, however, warmer and have a weaker and more inclined field, consistent with their larger distance from the centre of the spot. 

The scatter plots of the different physical quantities of the standard filament, including its surroundings, show a qualitative similarity with the scatter plots of the same quantities for the full penumbra. In particular, the standard filament shows all the populations of points that are also visible in scatter plots of the whole penumbra. This confirms that the spines, together with the penumbral filaments, are the fundamental constituents of the sunspot penumbra. In particular, this similarity suggests that there are no further major components of the penumbra besides these two.

\begin{acknowledgements}
We would like to thank the referee for constructive comments. SKT would like to thank Ron Moore, Allen Gary, David Hathaway, and Mitzi Adams for their useful comments and/or discussion on this work. Hinode is a Japanese mission developed and launched by ISAS/JAXA, collaborating with NAOJ as a domestic partner, NASA and STFC (UK), as international partners. Scientific operation of the Hinode mission is conducted by the Hinode science team organised at ISAS/JAXA. This team mainly consists of scientists from institutes in partner countries. Support for the post-launch operation is provided by JAXA and NAOJ (Japan), STFC (U.K.), NASA (U.S.A.), ESA, and NSC (Norway). This work has been partially supported by the BK21 Plus Program through the National Research Foundation (NRF), funded by the Ministry of Education of Korea. SKT is supported by an appointment to the NASA Postdoctoral Program at the NASA Marshall Space Flight Center, administered by Oak Ridge Associated Universities through a contract with NASA. This research has made extensive use of NASA's Astrophysics Data System (ADS).
\end{acknowledgements}


\end{document}